\documentclass[useAMS,usenatbib]{mn2e}

\usepackage{graphicx,color,afterpage}
\usepackage[total={17.8cm,24.0cm},centering]{geometry}
\usepackage{times}
\usepackage{subfigure}
\newcommand{\hi}{{\sc H\,i}}
\def\msun{$M_{\odot}$}
\def\mum{$\mu$m}
\def\sauron{{\tt SAURON}}

\def\atlas{{{ATLAS}}$^{\rm 3D}$}
\def\kms{km s$^{-1}$}

\def\arcsec{$^{\prime \prime}$}
\definecolor{Mygrey}{gray}{0.75}

\newcommand{\ltsimeq}{\raisebox{-0.6ex}{$\,\stackrel{\raisebox{-.2ex}{$\textstyle <$}}{\sim}\,$}}

\newcommand{\farc}{\mbox{\ensuremath{.\!\!^{\prime\prime}}}}

\usepackage[compact]{titlesec}
\titlespacing{\section}{0pt}{*2}{*1}

\title[Star formation suppression in ETGs]{The \atlas\ Project - XXVIII. Dynamically-driven star formation suppression in early-type galaxies} 
\author[Timothy A. Davis et al.]{\parbox{\textwidth}{Timothy A. Davis$^{1}$\thanks{E-mail:\texttt{tdavis@eso.org}}, 
Lisa M. Young$^{2,3}$, 
Alison F. Crocker$^{4}$,
Martin Bureau$^{5}$,
Leo Blitz$^{6}$,
Katherine Alatalo$^{7}$,
Eric Emsellem$^{1,8}$,
Thorsten Naab$^{9}$,
Estelle Bayet$^{5}$,
Maxime Bois$^{10}$,
Fr\'ed\'eric Bournaud$^{11}$,
Michele Cappellari$^{5}$,
Roger L. Davies$^{5}$,
P. T. de Zeeuw$^{1,12}$,
Pierre-Alain Duc$^{11}$,
Sadegh Khochfar$^{13}$,
Davor Krajnovi\'c$^{14}$,
Harald Kuntschner$^{15}$,
Richard M. McDermid$^{16}$,
Raffaella Morganti$^{17,18}$,
Tom Oosterloo$^{17,18}$,
Marc Sarzi$^{19}$,
Nicholas Scott$^{20}$,
Paolo Serra$^{17,21}$,
and Anne-Marie Weijmans$^{22}$}\vspace{0.4cm}\\ 
\parbox{\textwidth}{
$^{1}$European Southern Observatory, Karl-Schwarzschild-Str. 2, 85748 Garching, Germany\\
$^{2}$Physics Department, New Mexico Institute of Mining and Technology, Socorro, NM 87801, USA\\
$^{3}$Academia Sinica Institute of Astronomy \& Astrophysics, PO Box 23-141, Taipei 10617, Taiwan, R.O.C.\\
$^{4}$Ritter Astrophysical Observatory, University of Toledo, Toledo, OH 43606, USA\\
$^{5}$Sub-Dept. of Astrophysics, Dept. of Physics, University of Oxford, Denys Wilkinson Building, Keble Road, Oxford, OX1 3RH, UK\\
$^{6}$Department of Astronomy, Campbell Hall, University of California, Berkeley, CA 94720, USA\\
$^{7}$ Infrared Processing and Analysis Center, California Institute of Technology, Pasadena, California 91125, USA \\
$^{8}$Universit\'e Lyon 1, Observatoire de Lyon, Centre de Recherche Astrophysique de Lyon and Ecole Normale Sup\'erieure de Lyon, 9 avenue Charles Andr\'e, F-69230 Saint-Genis Laval, France\\
$^{9}$Max-Planck-Institut f\"ur Astrophysik, Karl-Schwarzschild-Str. 1, 85741 Garching, Germany\\
$^{10}$Observatoire de Paris, LERMA and CNRS, 61 Av. de l`Observatoire, F-75014 Paris, France\\
$^{11}$Laboratoire AIM Paris-Saclay, CEA/IRFU/SAp -- CNRS -- Universit\'e Paris Diderot, 91191 Gif-sur-Yvette Cedex, France\\
$^{12}$Sterrewacht Leiden, Leiden University, Postbus 9513, 2300 RA Leiden, the Netherlands\\
$^{13}$Royal Observatory Edinburgh, Blackford Hill, Edinburgh, EH9 3HJ, UK\\
$^{14}$Leibniz-Institut f\"ur Astrophysik Potsdam (AIP), An der Sternwarte 16, D-14482 Potsdam, Germany\\
$^{15}$Space Telescope European Coordinating Facility, European Southern Observatory, Karl-Schwarzschild-Str. 2, 85748 Garching, Germany\\
$^{16}$Australian Astronomical Observatory, PO Box 915, North Ryde, NSW 1670, Australia \\
$^{17}$Netherlands Institute for Radio Astronomy (ASTRON), Postbus 2, 7990 AA Dwingeloo, The Netherlands\\
$^{18}$Kapteyn Astronomical Institute, University of Groningen, Postbus 800, 9700 AV Groningen, The Netherlands\\
$^{19}$Centre for Astrophysics Research, University of Hertfordshire, Hatfield, Herts AL1 9AB, UK\\
$^{20}$Sydney Institute for Astronomy (SIfA), School of Physics, The University of Sydney, NSW 2006, Australia\\
$^{21}$CSIRO Astronomy \& Space Science, PO Box 76, Epping, NSW 1710, Australia \\
$^{22}$School of Physics and Astronomy, University of St Andrews, North Haugh, St Andrews KY16 9SS, UK}}

\begin{document}

\date{Accepted 2014 March 19.  Received 2014 March 18; in original form 2013 December 18}

\pagerange{\pageref{firstpage}--\pageref{lastpage}} \pubyear{2012}

\maketitle

\label{firstpage}

\begin{abstract}
We present measurements of the star formation rate (SFR) in the early-type galaxies (ETGs) of the \atlas\ sample, based on \textit{Wide-field Infrared Survey Explorer} (\textit{WISE}) 22~\mum\ and \textit{Galaxy Evolution Explorer} far-ultraviolet emission. We combine these with gas masses estimated from $^{12}$CO and \hi\ data in order to investigate the star formation efficiency (SFE) in a larger sample of ETGs than previously available. We first recalibrate (based on \textit{WISE} data) the relation between old stellar populations (traced at $K_{\rm s}$-band) and 22~\mum\ luminosity, allowing us to remove the contribution of 22~\mum\ emission from circumstellar dust. We then go on to investigate the position of ETGs on the Kennicutt-Schmidt (KS) relation. Molecular gas-rich ETGs have comparable star formation surface densities to normal spiral galaxy centres, but they lie systematically offset from the KS relation, having lower star formation efficiencies by a factor of $\approx$2.5 (in agreement with other authors). This effect is driven by galaxies where a substantial fraction of the molecular material is in the rising part of the rotation curve, and shear is high. We show here for the first time that although the number of stars formed per unit gas mass per unit time is lower in ETGs, it seems that the amount of stars formed per free-fall time is approximately constant. The scatter around this dynamical relation still correlates with galaxy properties such as the shape of the potential in the inner regions. This leads us to suggest that dynamical properties (such as shear or the global stability of the gas) may be important second parameters that regulate star formation and cause much of the scatter around star-formation relations. 
\end{abstract}


\section{Introduction}

Star formation is a fundamental process, responsible for converting the soup of primordial elements present after the big bang into the universe we see around us today. Despite this, debate still rages about the way star formation proceeds, and the role (if any) that environment plays in its regulation.
For instance, high-redshift starbursts seem to convert gas into stars much more efficiently than local disc galaxies \citep{2010ApJ...714L.118D,2010MNRAS.407.2091G}. This increased efficiency may be explained by a change in gas properties (e.g the high fraction of gas at high volume densities in starbursts), or may be an artefact of the imperfect methods we have of estimating star-formation rates, and tracing molecular hydrogen \citep{2012ApJ...746...69G}.

Atomic gas is present in $\approx$32\% of early-type galaxies  \citep[ETGs;][]{1977A&A....54..641B,1985AJ.....90..454K,2006MNRAS.371..157M,2007A&A...474..851D,2009A&A...498..407G,2010MNRAS.409..500O,2012MNRAS.422.1835S}, dust in $\approx$60\% \citep{2001AJ....121..808C,2012ApJ...748..123S,2013MNRAS.431.1929A}, and molecular gas in 22\% \citep[][hereafter Paper IV]{2007MNRAS.377.1795C,Welch:2010in,2011MNRAS.414..940Y}. Low level residual star formation has also been detected through studies of UV emission \citep[e.g.][]{2005ApJ...619L.111Y,2007ApJS..173..619K,2010ApJ...714L.290S,Wei:2010bt}, optical emission lines \citep[e.g.][]{Crocker:2011ic} and infra-red emission (e.g. \citealt{1989ApJS...70..329K,2007MNRAS.377.1795C,2009ApJ...695....1T}, hearafter T09; \citealt{2010MNRAS.402.2140S}).

Typically ETGs have much smaller fraction of molecular gas to stellar mass than spirals. This average fraction appears to decrease with increasing galaxy bulge fraction (\citealt{2013MNRAS.432.1862C}, hereafter Paper XX; see also \citealt{2012ApJ...758...73S}). This suggests a connection between bulge formation and galaxy quenching, as also suggested by optical studies \citep{2012ApJ...753..167B}. However the decrease of the molecular gas fraction does not seem to be the only factor making ETGs red. In fact, even at fixed gas fraction, molecule-rich ETGs form stars less efficiently than normal spirals, and very much less efficiently than high-redshift starburst galaxies \citep[][hereafter Paper XXII]{2011MNRAS.415...61S,2012ApJ...758...73S,2013MNRAS.432.1914M}. Such a suppression would help explain how objects in the red sequence can harbour substantial cold gas reservoirs for a long period of time, without becoming significantly blue. 
A similar suppression of star formation may also be ongoing in the central parts of our own Milky Way \citep{2013MNRAS.429..987L}, suggesting this may be a general process in spheroids and/or dense stellar environments.
The physics of whatever process is causing this suppression of star formation is, however, unknown. The deep potential wells of these objects could hold gas stable against collapse \citep[dubbed `morphological-quenching';][]{2009ApJ...707..250M}, or strong tidal fields and streaming motions could pull clouds apart \citep[e.g.][]{2013arXiv1304.7910M,2013arXiv1303.6286K}, lowering the observed SFE.

In this work we use data from the \atlas\ project to investigate if local ETGs do display a lower SFE than local spirals, and if so what may be driving this suppression. \atlas\ is a complete, volume-limited exploration of local ($<$42 Mpc) ETGs \citep[][hereafter Paper I]{2011MNRAS.413..813C}. All 260 \atlas\ sample galaxies have measured total molecular gas masses (or upper limits; from IRAM 30m CO observations presented in Paper IV). \hi\ masses are also available for the northern targets (from Westerbork Synthesis Radio Telescope, WSRT, observations; \citealt{2012MNRAS.422.1835S}, hereafter Paper XIII). To estimate the SFR in these objects, we utilise data from the \textit{Wide-field Infrared Survey Explorer} \citep[\textit{WISE};][]{2010AJ....140.1868W} all sky survey at 22~\mum, and from the \textit{Galaxy Evolution Explorer} (\textit{GALEX}) in the far ultraviolet (FUV).

Section 2 presents the data we use in this work, and describes how derived quantities are calculated. Section 3 presents our results, where we investigate the 22~\mum\ emission from CO non-detected ETGs, and the star formation activity in objects with a cold ISM. Section 4 discusses these results, and what we can learn about star formation and the evolution of ETGs. Section 5 presents our conclusions.

\section{Data}
\label{data}
In this paper we consider the \atlas\ sample of ETGs. This sample was carefully selected based on morphology to include every early-type object (brighter than -21.5 in $K_{\rm s}$-band) visible from the William Herschel Telescope, out to a distance of 42 Mpc, and is thus a complete, volume-limited sample. More information on the sample selection can be found in Paper I. In this work we consider two sub-samples, those galaxies with a detected molecular ISM that can provide fuel for star formation (from Paper IV), and those without. Here we consider entire galaxies in an integrated manner. A spatially-resolved star-formation analysis will be presented in a future work. To estimate the star formation efficiency in these objects, we require both molecular and atomic gas masses, tracers of obscured and un-obscured star formation, and sizes for the regions concerned. We describe below how these were obtained.

\subsection{Molecular gas masses}

The CO(1-0) and CO(2-1) lines were observed in every galaxy in the \atlas\ sample at the IRAM-30m telescope, and 56 objects were detected (for full details see Paper IV). From these observations we have estimated molecular gas masses for the detected galaxies, using a Galactic $X_{\rm CO}$ factor of 3$\times10^{20}$ cm$^{-2}$ (K \kms)$^{-1}$ \citep{Dickman:1986jz}. We return to discuss this assumption later, but as ETGs usually have high stellar metallicities such a value is a priori reasonable. Making this assumption, we found molecular gas reservoirs with masses between 10$^6$ and 10$^{9.5}$ \msun, as tabulated in Paper IV.
We were also able to place limits on the amount of molecular gas of CO non-detected objects, finding upper limits between 10$^6$ and 10$^8$ \msun (for objects at different distances).  

These observations were single pointings at the galaxy centres, with a beam size of $\approx$22\arcsec\ for the CO(1-0) transition (used to calculate the molecular gas masses). In some objects the molecular gas distribution was later shown to be more extended than the 30m telescope beam (see \citealt{2013MNRAS.429..534D}, hereafter Paper XIV, for an analysis of the total molecular gas extent in these objects). In these cases, we use total interferometric CO fluxes from \citet[][hereafter Paper XVIII]{2013MNRAS.432.1796A}. In principle it is possible that these interferometric observations resolved out some emission, which would make our CO masses lower limits. The correction for molecular material outside the beam of our single-dish observations is much more significant, however, and so we consider it better to use the interferometric fluxes where possible. As the CO is not generally extremely extended, we do not expect the amount of flux resolved out to be large, so this should not affect our conclusions.  In objects without interferometric observations, we used the single-dish CO fluxes to estimate the masses. Our size estimates (described below) suggest that very few of these unmapped objects have extended gas reservoirs, so these 30m telescope measurements are unlikely to miss substantial amounts of molecular material. 

\subsection{Atomic gas masses}

As presented in Paper XIII, all \atlas\ field galaxies above a declination of 10$^{\circ}$ were observed with the WSRT, with a resolution of $\approx$35\arcsec. For Virgo cluster objects we take the data from the ALFALFA survey \citep{2007A&A...474..851D}, as documented in Paper XIII. Most of the molecular discs studied here are smaller than 35\arcsec, so we assume that only the \hi\ gas mass detected in the innermost beam is important. The central \hi\ mass used here is listed in Table A1 of Young et al (2013). In many cases the \hi\ in the central regions is unresolved. When calculating the combined gas surface density we assume that the \hi\ is cospatial with the CO. In galaxies with large \hi\ discs (Class "D" in Paper XIII), we assume that the \hi\ disc has a uniform surface density over the entire inner beam. This is an assumption, but has been shown to be reasonable in other galaxies, where \hi\ emission saturates in the inner parts of the discs \citep[e.g.][]{2002ApJ...569..157W,2008AJ....136.2846B}. In objects below a declination of 10$^{\circ}$, where we have no \hi\ observations, we assume that the \hi\ mass is negligible. We expect this assumption to be reasonable, given that the other objects we study here are all molecular-gas dominated in the inner parts.

\subsection{22~\mum\ fluxes}
\label{22umfluxes}
Emission at $\approx$20-25 \mum\ traces warm dust, that is present around hot newly-formed stars, in the ejected circumstellar material around hot old stars, and in AGN torii. If one can correct for the emission from old stars (in the absence of strong AGN), the $\approx$20-25 \mum\ emission can provide a sensitive estimate of the amount of obscured star formation in our systems.

Here we use 22~\mum\ fluxes from the \textit{WISE} catalogue \citep{2010AJ....140.1868W} all sky data release. We chose to use \textit{WISE} 22~\mum\ rather than \textit{Spitzer} 24 \mum\ observations as the \textit{WISE} data are available for every source in our sample, at a reasonably uniform depth (and \citealt{2014arXiv1402.3597C} have shown that where multiple measurements exist the scatter between \textit{Spitzer} 24 \mum\ and \textit{WISE} 22 \mum\ fluxes is low). 
We downloaded the \textit{WISE} 22~\mum\ catalogue profile fit magnitudes (\textit{w4mpro}) and aperture magnitude values (parameter \textit{w4gmag}) from the \textit{WISE} catalogue \citep{2010AJ....140.1868W}. The aperture values are calculated using elliptical apertures defined from the position, size and inclination of the galaxy from the Two Micron All Sky Survey \citep[2MASS][]{Skrutskie:2006p2829} Extended Source Catalog (XSC; \citealt{Jarrett:2000p2407}), and enlarged by the \textit{WISE} team to correct for the larger point-spread function of the \textit{WISE} satellite. See the \textit{WISE} documentation\footnote{http://wise2.ipac.caltech.edu/docs/release/allsky/ - accessed 30/05/13} for full details of these magnitudes. 

As some of our sources are (marginally) extended compared to the \textit{WISE} beam, {and the profile fit magnitudes are known to underestimate the true flux for extended sources}, we preferentially use the aperture flux values (parameter \textit{w4gmag}). In CO-detected objects we verified that the apertures used in the catalogue are always larger than the CO distribution. In a few objects (usually the most star-forming objects with compact gas reservoirs) the profile fit magnitudes retrieve more flux, and so we instead use these \textit{w4mpro} values. The 22~\mum\ fluxes we measure for each object (and the respective errors, as listed in the catalogue) are listed in Table \ref{obstable} for our CO detected sample. For the CO non-detected sample we always use the aperture magnitudes, and list the derived 22~\mum\ luminosities in Table \ref{passivetable}. The method we use to calculate star formation rates while removing the contribution of old stars is discussed in Sections \ref{passive} and \ref{sfrs}.

\subsection{FUV fluxes}

FUV light is emitted primarily by young O and B stars, and hence traces star formation activity over the last $\approx$0.1 Gyr. In the most massive and old ETGs, the UV-upturn phenomenon is observed, in which additional FUV light is emitted by an older population of stars (likely post-asymptotic giant branch stars; \citealt{2008ASPC..392....3Y,2011MNRAS.414.1887B}). The light from this phenomenon is diffuse, following the old population, and is energetically unimportant if even low-level star formation is present \citep{2005ApJ...619L.111Y}. Thus we do not expect this phenomenon to substantially affect measurements of star formation derived from FUV in this work. If it were to have an effect, however, it would formally make our star formation estimates upper limits.

FUV magnitudes for the star-forming galaxy sample used in this work were obtained from the \textit{GALEX} catalog server, release GR7. Where multiple observations of the same target exist, we always used the deepest observation. These magnitudes are corrected for foreground extinction assuming the Milky Way E(B-V) values from \cite{1998ApJ...500..525S} scaled to UV using A$_{\rm FUV}$ = 8.24 E(B-V) \citep{2007ApJS..173..185G}. 
The FUV flux measured for each object (and its error) is listed in Table \ref{obstable}. The method used to calculate star formation rates is discussed in Section \ref{sfrs}.

\subsection{Source sizes}

To estimate the mean surface density of gas and star formation tracers, one needs to know the total area over which they are distributed. For most of the objects in our star-forming sample, the area can be directly estimated from the size of the molecular gas reservoir seen in our interferometric CO observations. This size is listed in Column 3 of Table 1 in Paper XIV\footnote{In the published version of Paper XIV the size entries for several galaxies were incorrect. We here use the correct $D_{CO}$ values of 20\arcsec for NGC4150 and 21\farc2 for NGC4526}. The typical gas reservoir is found to have a radial extent of $\approx$1 kpc. In Paper XIV we also present a `beam corrected' size for the CO reservoirs, but as such a process is intrinsically uncertain we here choose to use the observed measured extents (making our adopted sizes formally upper limits).
If at our resolution the gas appears to be in a disc, the area is estimated assuming the gas is in a flat circular disc, with a diameter given by the observed major-axis length. Where our observations reveal rings of gas (either spatially or in velocity space; see Paper XIV and Paper XVIII), the rings are assumed to have a radial width of 200 pc. This is an assumption based purely on the size of the rings that are visible in optical images in some well-resolved cases (e.g. NGC~4324). If the molecular rings were smaller, the surface density of gas used would be underestimated. The galaxies in which we make this correction are NGC~2685, NGC~2764, NGC~3626, NGC~4324 and NGC~5866. 

For those objects where only single-dish molecular gas data exist, we estimate the area of the star-forming regions using resolved images of gas and/or star-formation tracers. In this work we estimate the size of these regions using the highest resolution tracer available. Where possible we use \textit{Hubble Space Telescope} (\textit{HST}) images of UV emission, or unsharp-masked optical \textit{HST} images that pick out patchy dust (that has been shown to be almost always cospatial with the cold gas, occasionally slightly more extended; Paper XVIII). Where \textit{HST} observations are not available, we use the size of the FUV-emitting region, as estimated from \textit{GALEX} images (resolution $\approx$6\arcsec), the size of strong Balmer line emitting regions in our \sauron\ integral field unit observations (resolution $\approx$1-2\arcsec), or the size of the 24 \mum\ emission in \textit{Spitzer} images (resolution $\approx$6\arcsec). The source sizes we measure, and the data these are based on, are listed in Table \ref{obstable}. We include the additional uncertainty in size coming from the limited spatial resolution of some of these data, as shown in Table \ref{obstable}. We use these source sizes to estimating the surface densities of both star formation and gas tracers.

\section{Results}

\subsection{22~\mum\ emission from CO non-detected ETGs}
\label{passive}
  \begin{figure}
\begin{center}
\includegraphics[width=0.5\textwidth,angle=0,clip,trim=0.0cm 0cm 0cm 0cm]{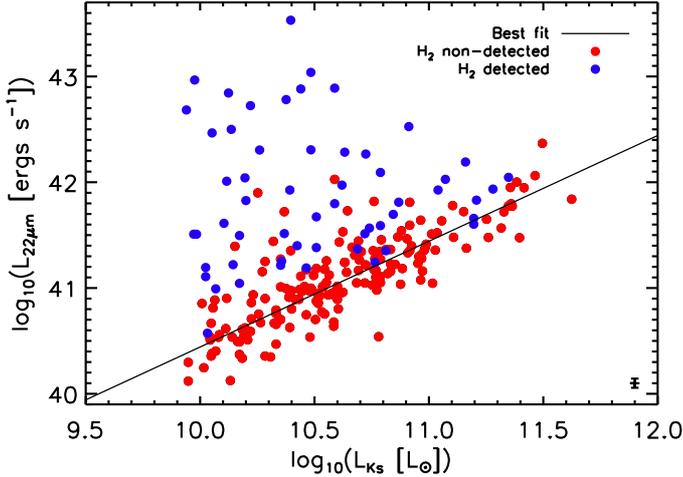}
 \end{center}
 \caption{22 $\mu$m \textit{WISE} luminosities of the \atlas\ galaxies plotted against their $K_{\rm s}$-band luminosities. Blue circles are galaxies with detected molecular gas; red circles are those ETGs without a detected molecular ISM. The best-fit to the \atlas\ galaxies without a detected molecular ISM is shown as a black solid line. The typical error on each point is shown in the bottom-right corner of the plot.}
 \label{WISEpassive}
 \end{figure}

As discussed above, the hot dust that gives rise to 22~\mum\ emission from galaxies occurs both in the birth clouds around newly formed massive stars and in the circumstellar ejecta of old stars. T09 studied the 24 \mum\ emission of 18 CO non-detected elliptical galaxies from the \sauron\ galaxy sample (\citealt{deZeeuw:2002gq}; a subset of the sample studied here). They found that the 24 \mum\ emission from these objects correlates well with the $K_{\rm s}$-band luminosity (a proxy for stellar mass), as would be expected from emission from an old stellar population.

We here reproduce such a correlation in Figure \ref{WISEpassive}, but using 22~\mum\ \textit{WISE} luminosities for all 260 galaxies of the \atlas\ sample. 
Our sample galaxies that contain no detectable molecular ISM are shown as red circles, while molecular gas-rich objects are shown in blue. A typical error bar is shown in the bottom-right corner of the plot.
The $K_{\rm s}$-band luminosity of each object has been estimated from its 2MASS $K_{\rm s}$-band magnitude, assuming that the absolute magnitude of the Sun at $K_{\rm s}$-band is 3.28 mag \citep[Table 2.1 of][]{Binney:1998p3454}. To be consistent with the other papers in this series, we use the $K_{\rm s,total}$ magnitude (parameter  k\_m\_ext from the 2MASS catalogue; \citealt{Jarrett:2000p2407,Skrutskie:2006p2829}), as tabulated in Paper I.  These $K_{\rm s,total}$ magnitudes are measured over large apertures, to include the total flux from the galaxy using the techniques developed in \cite{Kron:1980p3008} and curves-of-growth (see \citealt{Jarrett:2000p2407} for further details). Distances to these galaxies are given in Paper I. The \textit{WISE} and 2MASS luminosities we derived for the sample objects are listed in Table \ref{passivetable}. 

\begin{figure*}
\begin{center}
\includegraphics[width=0.9\textwidth,angle=0,clip,trim=0.0cm 0cm 0cm 0cm]{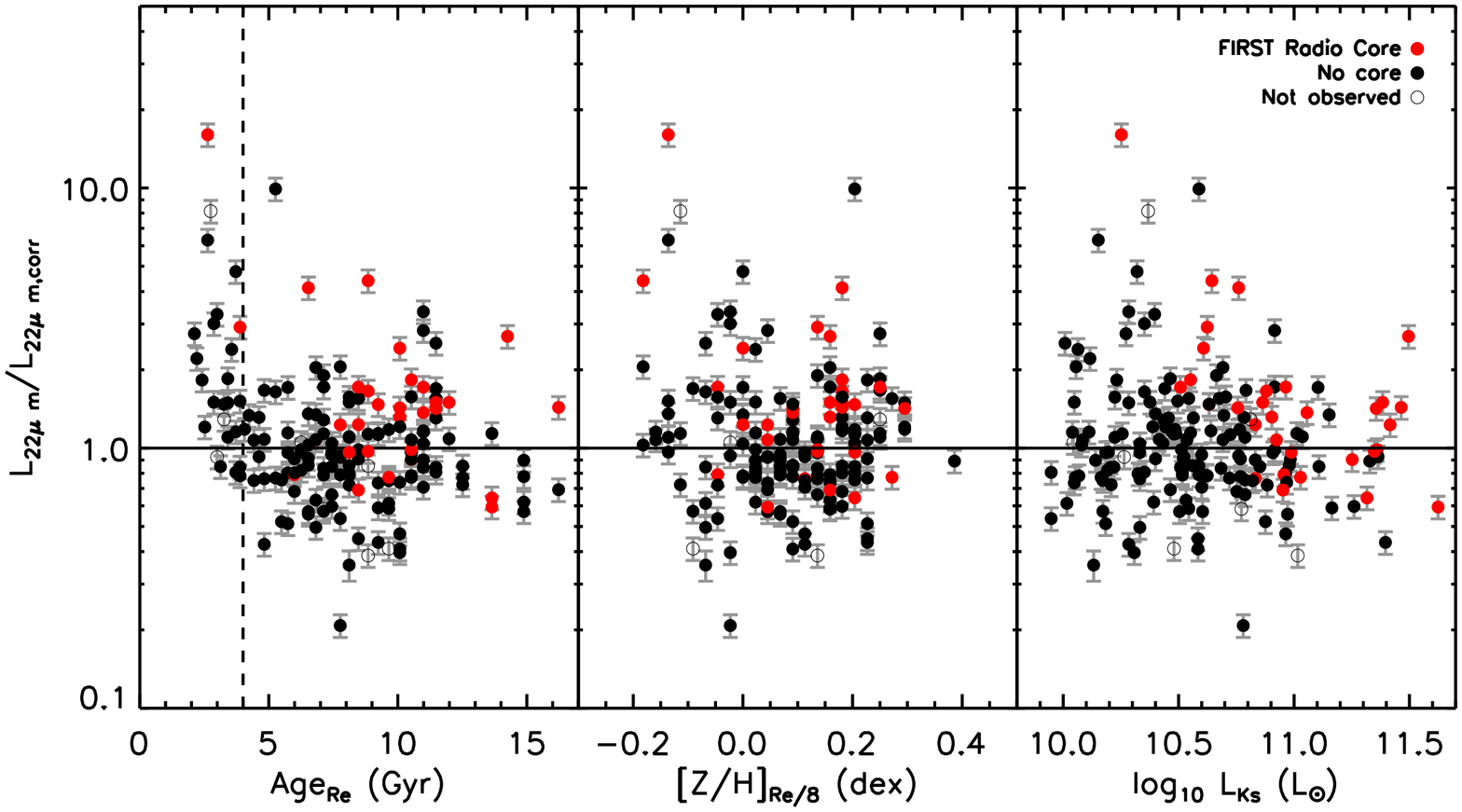}
 \end{center}
 \caption{Residuals around the best-fit line from Figure \protect \ref{WISEpassive} plotted as a function of galaxy properties, for the CO non-detected \atlas\ ETGs. The left panel shows residuals versus the age of the stellar population in the galaxy, measured within a one effective radius aperture (McDermid et al., 2013). The vertical dashed line in this panel is a guide to the eye at a population age of 4 Gyr (as discussed in the text). In the centre panel the residuals are plotted as a function of the central metallicity of the galaxy (measured in an Re/8 aperture; using aperture values with one effective radius would not change our conclusions). The right panel shows the residuals against the $K_{\rm s}$-band luminosity of the galaxy, as measured for Figure \protect \ref{WISEpassive}. The solid black line is our best-fit relation from Figure \protect \ref{WISEpassive}.   Red circles denote galaxies which have a compact radio core, and black circles those without. Open circles show objects that are not in the FIRST survey volume.}
 \label{passive_dependances}
 \end{figure*}

Our CO non-detected galaxy sample shows a clear correlation with galaxy luminosity, but with a significant scatter. Galaxies with molecular gas show no clear correlation between their 22~\mum\ emission and stellar luminosity, but always lie above the mean location of the CO non-detected galaxies for any given stellar luminosity, strengthening our suspicion that the bulk of their 22~\mum\ emission is star-formation related. Some galaxies in our CO non-detected sample ($\approx$10\%) also lie well above the relation formed by the majority of the CO non-detections. {Some these galaxies lie systematically at the edge of our survey volume, where our molecular gas detection limit is highest, and are thus likely star-forming systems which lie below our CO detection limit. Others have young stellar population ages in their central parts (suggesting they may have been star-forming in the recent past) or have active AGN. We discuss these objects in more detail in Figure \ref{passive_dependances} and below}.

We conduct a robust fit (using the LTS\_LINEFIT\footnote{available at http://www-astro.physics.ox.ac.uk/~mxc/idl/\#lts} routine described in \citealt{2013MNRAS.432.1709C}; hereafter Paper XV) to determine the mean amount of 22~\mum\ emission caused by old stellar populations at each stellar mass. The coefficients of the best fit relation are shown in Equation \ref{my_passivecorr}, and the systems that were considered outliers are indicated with a cross in Table \ref{passivetable}. We note that doing a simple fit including all the outliers (that are likely star-forming) would slightly change the slope of the derived relation, but would not alter our conclusions. 

\begin{equation}
\log \left(\frac{L_{\rm 22\mu m, passive}}{\rm ergs\,s^{-1}}\right)=(1.00\pm0.04) \log\left(\frac{L_{K_s}}{L_{\odot}}\right) + (30.45\pm0.46)
\label{my_passivecorr}
\end{equation}

In order to estimate the amount of 22~\mum\ emission arising exclusively from star formation in our CO detected sample, we subtract off the contribution of the passive stellar populations (following T09):

\begin{equation}
L_{\rm 22\mu m, SF}=L_{\rm 22\mu m, obs} - L_{\rm 22\mu m, passive},
\label{my_passivecorr2}
\end{equation}

\noindent where L$_{\rm 22\mu m, passive}$ is obtained from the $K_{\rm s,total}$ luminosity via Equation \ref{my_passivecorr}.

For CO non-detected galaxies, the scatter around the best-fit relation in figure \ref{WISEpassive} is large ($\approx$0.4 dex), larger than the expected uncertainty in either luminosity. We searched for an astrophysical explanation for this intrinsic scatter.
Galaxies with no detected molecular ISM that have large \hi\ discs, clouds or disturbed \hi\ distributions do not show any enhancement in 22~\mum\ emission over and above that expected for a passive population. Galaxies with small \hi\ discs do lie above our best-fit relation for CO non-detected objects, consistent with having some small but non-negligible contribution from star formation at 22 \mum, but as only two cases are present in our sample these objects do not drive the intrinsic scatter observed.

When controlling for stellar luminosity, the offsets above and below the line defined in Equation \ref{my_passivecorr} for the CO non-detected sample do not correlate with stellar kinematic quantities \citep{2011MNRAS.414.2923K,2011MNRAS.414..888E}, ionised-gas quantities, or measures of galaxy environment \citep{2011MNRAS.416.1680C}. Stellar population age (or equivalently the strength of H$\beta$ absorption; McDermid et al., 2013) does show a weak trend (Figure \ref{passive_dependances}, left panel), in that the systems with the youngest ($<$4 Gyr) mean stellar populations (detected in any aperture) tend to lie above the best-fit relation (likely due to a larger number of AGB stars, that are important sources of dust creation from a stellar population). However the vast majority of galaxies in our CO non-detected sample are dominated by older stellar populations, and the residuals around Equation \ref{my_passivecorr} do not correlate with age beyond 4 Gyr.
\cite{2013ApJ...768...28M} found that the metallicity of the stellar population is an important driver of the scatter in this relation at fixed mass. With a larger sample of objects we are unable to reproduce this trend (Figure \ref{passive_dependances}, centre panel).

  Our CO non-detected galaxy sample does not contain many strong active galactic nuclei (AGN), but lower luminosity nuclear activity could contribute to the scatter seen in Figure \ref{WISEpassive} (as the torus region of an AGN emits in the mid-infrared; e.g. \citealt{2010ApJ...718.1171R}). The 31 galaxies in our CO non-detected sample that have radio cores in the Faint Images of the Radio Sky at Twenty-Centimeters (FIRST) survey \citep{1995ApJ...450..559B} do tend to lie above our best fit-relation (see Figure \ref{passive_dependances}, right panel). Removing galaxies with radio cores does not substantially affect our best-fit (Equation \ref{my_passivecorr}). Almost all the X-ray-bright AGN identified in our sample (by \citealt{2013MNRAS.432.1845S}) also have a molecular ISM, so they do not contribute to the scatter discussed here. The presence of central ionised-gas velocity dispersion peaks often correlates with low level nuclear activity (Sarzi et al., in prep), but we do not see any clear trend in the residuals of galaxies with such an enhancement.

\subsection{Star-formation rates}
\label{sfrs}

  \begin{figure}
\begin{center}
\includegraphics[width=0.45\textwidth,angle=0,clip,trim=0.0cm 0.0cm 0cm 0cm]{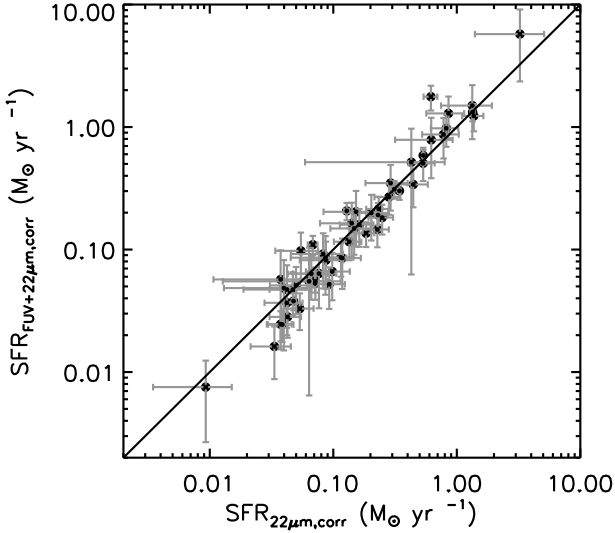}
 \end{center}
 \caption{Comparison of SFRs derived using \textit{WISE} 22~\mum\ emission only, and \textit{WISE} 22~\mum\  combined with \textit{GALEX} FUV emission for our star-forming galaxy sample. The solid line shows the 1:1 relation to guide the eye.}
 \label{WISE_galex_compfig}
 \end{figure}

Many different conversions exist to convert an observed flux in a given waveband to a SFR. These are primarily empirical conversions, often based on observations of nearby galaxies. 
Consistency between SFR estimates made using different tracers is thus only likely if the conversions are calibrated against the same set of sample galaxies. 
Here we use the results of \cite{2007ApJ...666..870C} to estimate SFRs from our measured 22~\mum\ fluxes. 
We also use a combined relation from \cite{2011ApJ...741..124H} to estimate the SFR from \textit{WISE} and \textit{GALEX} data together. {This combined relation extinction corrects the \textit{GALEX} fluxes, using the total infrared to FUV luminosity ratio (IRX) method. This extinction correction is very important, as without it FUV SFRs can be underestimated by half an order of magnitude (see \citealt{2011ApJ...741..124H} for full details).}
This allows us to estimate the contribution from both obscured and unobscured star formation (based on a Kroupa initial mass function, IMF, burst of age 1 Gyr).
Both of these calibrations are formally for 24 \mum\ \textit{Spitzer} observations, but the bandpasses of the \textit{Spitzer} 24 \mum\ and \textit{WISE} 22 \mum\ filters  (and the SEDs of galaxies in this region) are sufficiently similar that the error induced by using \textit{WISE} 22~\mum\ measurements should be minimal. Importantly, the star formation calibrations we use are both based on the \textit{Spitzer} Infrared Nearby Galaxies Survey, and thus should be internally consistent. 
Using a SFR estimator derived specifically for \textit{WISE} 22~\mum\ data but not calibrated on the same galaxy sample, such as that by \cite{2012JApA...33..213S}, would not change the conclusions of this paper.

When using 22~\mum\ fluxes in either of these two conversions considered we first remove the emission from the passive old stellar populations (using Equations \ref{my_passivecorr} and \ref{my_passivecorr2}, as discussed above). One CO-detected object falls below the correlation in Equations \ref{my_passivecorr} and \ref{my_passivecorr2} (NGC2768), suggesting it has low amounts of obscured SF, and thus the 22~\mum\ band is dominated by emission from old stellar populations. We remove this object from our analysis of 22 \mum\ SFRs from this point on, but do include this object in the combined 22 \mum\ + FUV relations, by assuming its 22 \mum\ flux is zero (and hence all the star-formation is unobscured).

For each galaxy in our star-forming sample, we list both SFRs we estimate in Table \ref{obstable}. The errors in these SFRs are estimated through propagation of the uncertainties in the input quantities, and these are also listed in Table \ref{obstable}.
The SFRs estimated from the 22~\mum\ emission alone agree well within the errors with those estimated from FUV emission combined with 22~\mum\ emission (see Figure \ref{WISE_galex_compfig}). The ratio of the SFRs derived with and without the FUV does not show any correlation with galaxy mass, confirming that the UV-upturn phenomenon is not adversely affecting the UV-derived SFRs. The ratio of these two star-formation rates may weakly depend on the star-formation rate itself {(the best fit relation with a fixed intercept between these two indicators has a slope of 1.03$\pm$0.02)}, but more data would be needed to confirm if this low significance trend is real. Overall, the agreement between these SFRs suggests ETGs have ratios of obscured and unobscured star formation similar to those of spiral galaxies.

   \begin{figure*}
\begin{center}
\includegraphics[height=5.45cm,angle=0,clip,trim=0.0cm 1.6cm 0cm 0cm]{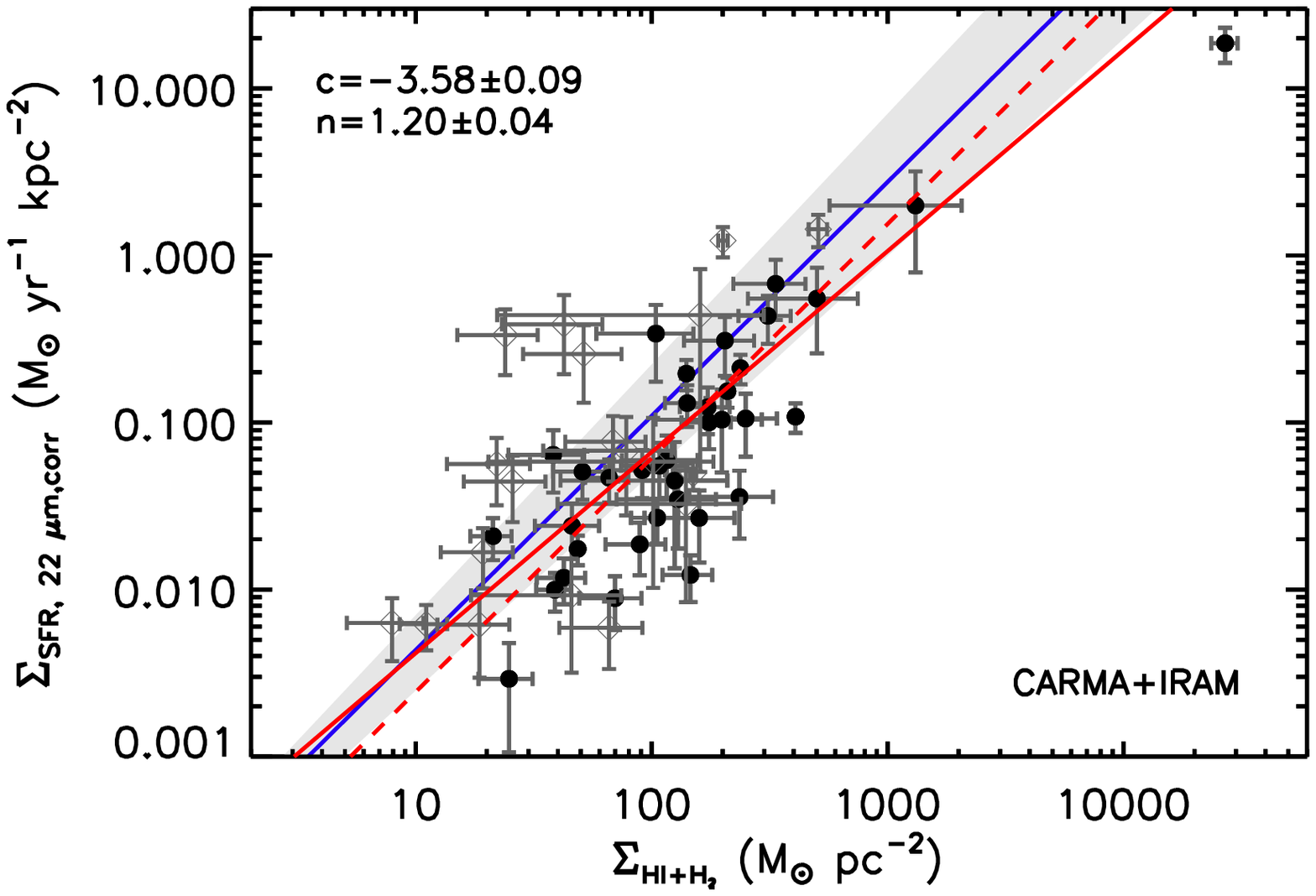}
\includegraphics[height=5.45cm,angle=0,clip,trim=3.1cm 1.6cm 0cm 0cm]{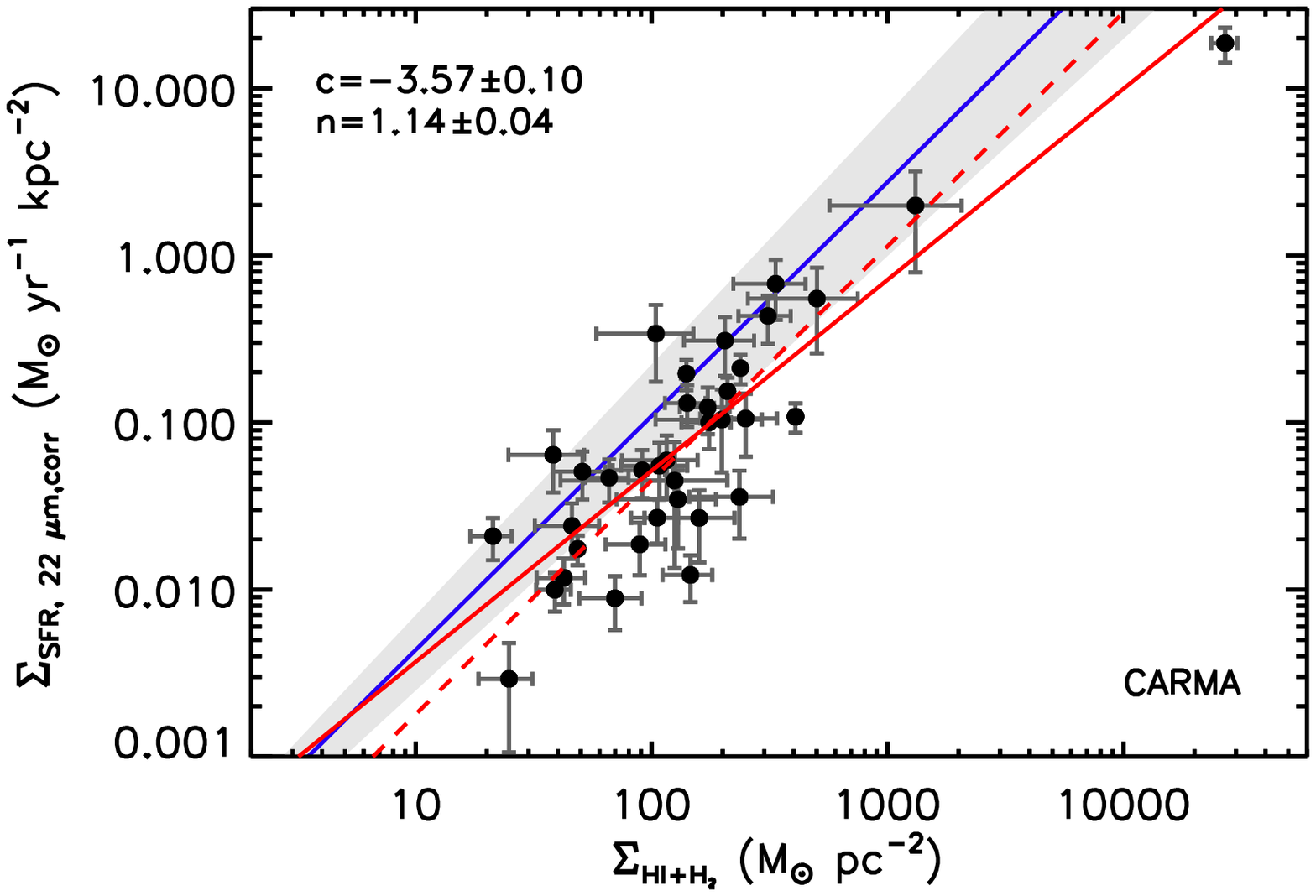}\\
\includegraphics[height=6.35cm,angle=0,clip,trim=0.0cm 0.0cm 0cm 0.0cm]{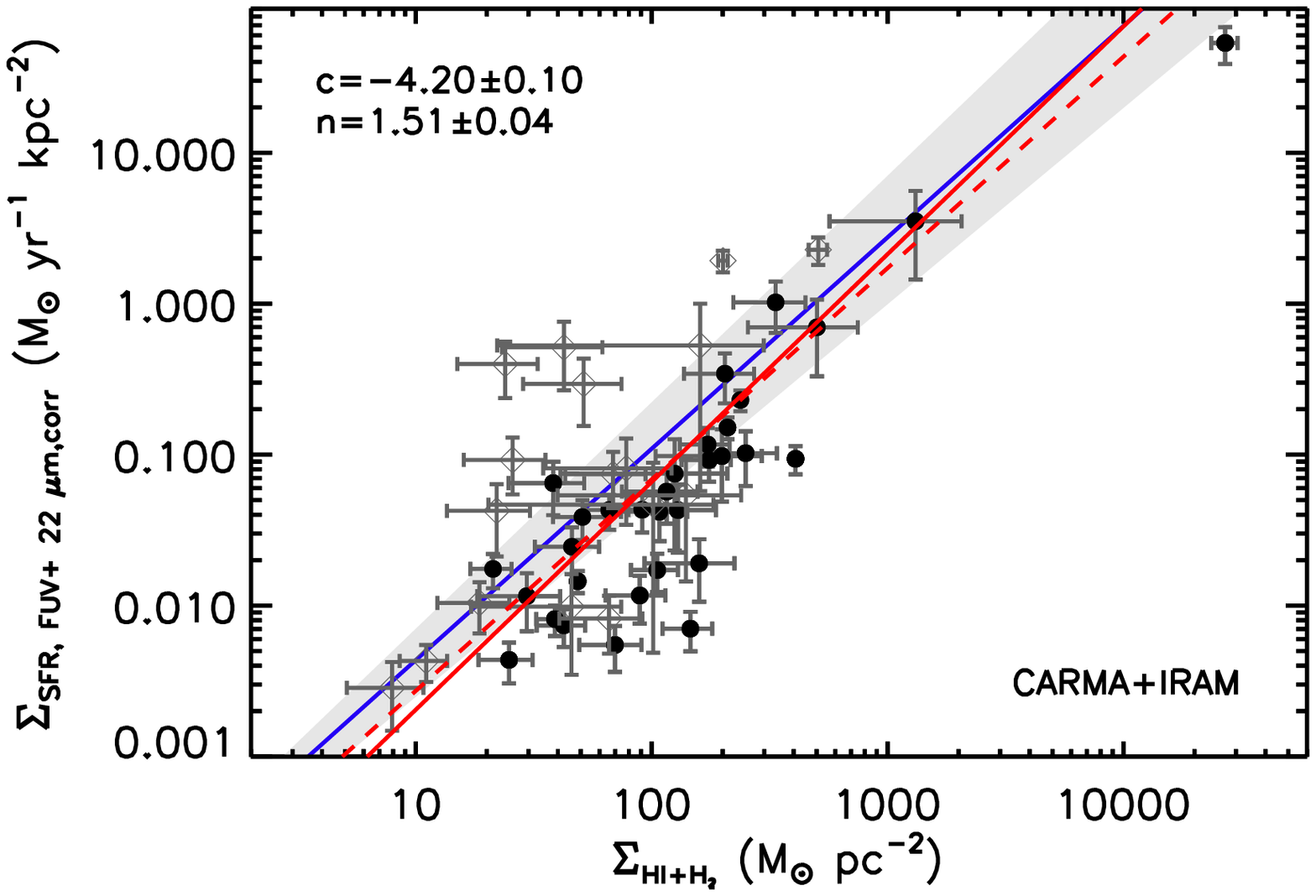}
\includegraphics[height=6.35cm,angle=0,clip,trim=3.1cm 0.0cm 0cm 0cm]{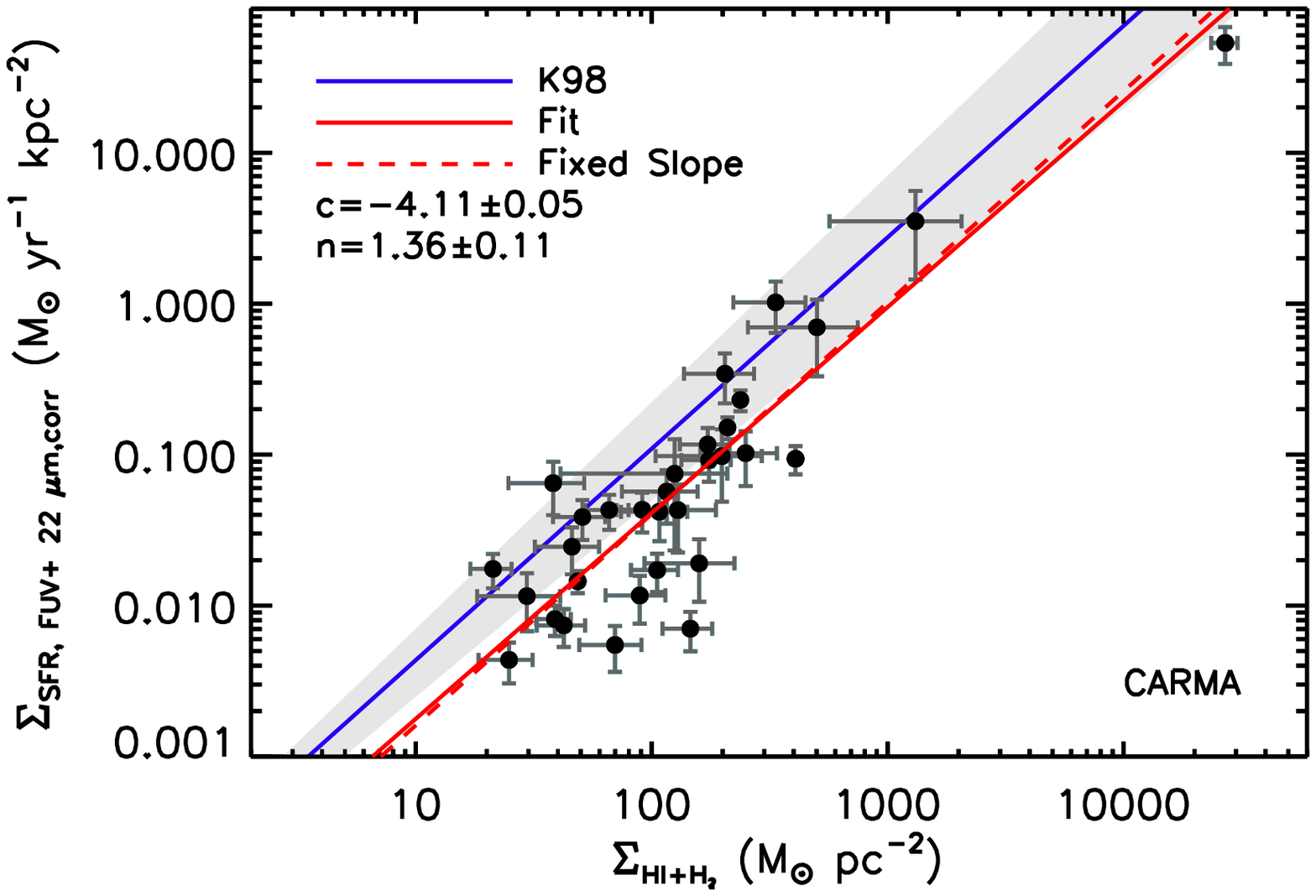}
 \end{center}
 \caption{The SFR surface density of molecular gas-rich \atlas\ ETGs, shown as a function of the gas (atomic and molecular) surface density. The H$_2$ + \hi\  surface densities are estimated as described in Section \ref{data}, and the SFRs as described in Section \ref{sfrs}. The data points have had the fraction of their 22~\mum\ emission arising from circumstellar emission removed, as described in Section \ref{passive}. The top row shows SFR densities derived from \textit{WISE} data only, while the bottom row shows SFRs derived from a combination of \textit{GALEX} FUV and \textit{WISE} 22~\mum\ emission. Black circles denote galaxies where resolved interferometry is available from Paper XVIII (allowing better estimates of the total molecular gas mass and density) and open symbols show galaxies for which only IRAM 30m telescope data are available. The left column shows all galaxies, while in the right column the IRAM points have been removed to include only our most reliably-determined data points. The galaxy which appears at the top right of every panel is NGC1266, which hosts a large molecular gas outflow (see \citealt{2011ApJ...735...88A,2012MNRAS.426.1574D} for more details). We show in all panels the K98 SF relation (for local spiral/starburst galaxies) converted to a Kroupa IMF as a solid blue line, with its typical scatter indicated as a grey shaded area. Our best-fit star-formation relations for ETGs are shown as red solid and dashed lines, for the relation with a free and fixed slope, respectively. The legend of each plot indicates the fitted slope and intercept of the best unconstrained fit.}
 \label{WISE22umfig}
 \end{figure*}

\begin{table*}
\caption{Properties of the star-forming ETG sample used in this work.}
\begin{tabular*}{1.00\textwidth}{@{\extracolsep{\fill}}l r r r r r r r r r r r r}
\hline
Galaxy & Area & Source & log $\Sigma_{\rm HI+H_2}$ & $\tau_{\rm dyn}$ & $\tau_{\rm ff}$ &  F$_{\rm 22\mu m}$ & F$_{\rm 22\mu m,corr}$ &  F$_{\rm FUV}$ & \scriptsize $\Sigma_{\rm SFR,22\mu m}$&\scriptsize  $\Sigma_{\rm SFR,FUV+22\mu m}$ \\
    &  (kpc$^2$) & & ($M_{\odot}$ pc$^{-2}$) & \multicolumn{2}{c}{(Myr)} &   (mJy) &  (mJy)   &  ($\mu$Jy) & \multicolumn{2}{c}{(log $M_{\odot}$ yr$^{-1}$ kpc$^{-2}$)}\\
     (1) & (2) & (3) & (4) & (5) & (6) & (7) & (8) & (9) & (10) & (11)\\
 \hline
IC0676 & 2.51 $\pm$ 0.83 & PXIV & 2.31 $\pm$ 0.22 & 66.1 & 8.1 & 550.0 & 545.5 $\pm$ \hspace{4pt}3.7 & 238 $\pm$ \hspace{4pt}4.1 & -0.51 $\pm$ 0.19 & -0.46 $\pm$ 0.18\\
IC0719 & 10.31 $\pm$ 2.04 & PXIV & 1.33 $\pm$ 0.17 & 51.5 & 14.3 & 58.7 & 53.9 $\pm$ \hspace{4pt}0.4 & 652 $\pm$ 27.6 & -1.68 $\pm$ 0.14 & -1.76 $\pm$ 0.12\\
IC1024 & 4.29 $\pm$ 1.02 & PXIV & 2.24 $\pm$ 0.19 & 63.2 & 8.4 & 339.1 & 336.0 $\pm$ \hspace{4pt}1.9 & 652 $\pm$ \hspace{4pt}8.6 & -0.91 $\pm$ 0.15 & -0.93 $\pm$ 0.14\\
NGC0509 & 3.81 $\pm$ 1.36 &  24$\mu$m & 0.90 $\pm$ 0.23 & 77.6 & 18.2 & 5.7 & 3.8 $\pm$ \hspace{4pt}0.2 & 2 $\pm$ \hspace{4pt}0.8 & -2.20 $\pm$ 0.20 & -2.54 $\pm$ 0.23\\
NGC0524 & 3.76 $\pm$ 0.97 & PXIV & 1.39 $\pm$ 0.19 & 18.3 & 13.7 & 51.5 & 2.8 $\pm$ \hspace{4pt}1.9 & 239 $\pm$ \hspace{4pt}7.5 & -2.53 $\pm$ 0.31 & -2.36 $\pm$ 0.15\\
NGC1222 & 1.63 $\pm$ 0.92 & PXIV & 3.12 $\pm$ 0.31 & 65.8 & 5.1 & 1824.6 & 1820.9 $\pm$ 17.7 & 2178 $\pm$ 17.8 & 0.30 $\pm$ 0.29 & 0.55 $\pm$ 0.29\\
NGC1266 & 0.03 $\pm$ 0.01 & A11 & 4.76 $\pm$ 0.16 & 16.8 & 2.0 & 734.4 & 728.9 $\pm$ \hspace{4pt}5.3 & 22 $\pm$ \hspace{4pt}4.2 & 1.27 $\pm$ 0.11 & 1.73 $\pm$ 0.13\\
NGC2685 & 0.93 $\pm$ 0.35 & PXIV & 1.66 $\pm$ 0.23 & 20.4 & 11.8 & 53.3 & 37.4 $\pm$ \hspace{4pt}0.9 & 2066 $\pm$ 10.3 & -1.35 $\pm$ 0.21 & -1.03 $\pm$ 0.20\\
NGC2764 & 8.63 $\pm$ 0.33 & PXIV & 2.46 $\pm$ 0.15 & 119.7 & 7.4 & 300.0 & 295.8 $\pm$ \hspace{4pt}1.8 & 473 $\pm$ 20.4 & -0.81 $\pm$ 0.10 & -0.82 $\pm$ 0.08\\
NGC2768 & 1.48 $\pm$ 0.57 & PXIV & 1.47 $\pm$ 0.24 & 12.9 & 13.1 & 51.7 & -3.9 $\pm$ \hspace{4pt}1.8 & 377 $\pm$ \hspace{4pt}9.0 & -  & -1.94 $\pm$ 0.20\\
NGC2824 & 7.80 $\pm$ 2.44 & PXIV & 2.03 $\pm$ 0.21 & 37.6 & 9.5 & 74.9 & 71.9 $\pm$ \hspace{4pt}0.4 & 78 $\pm$ \hspace{4pt}7.4 & -1.26 $\pm$ 0.18 & -1.38 $\pm$ 0.17\\
NGC3032 & 4.85 $\pm$ 1.01 & PXIV & 1.82 $\pm$ 0.18 & 56.4 & 10.7 & 136.9 & 132.1 $\pm$ \hspace{4pt}0.5 & 1124 $\pm$ \hspace{4pt}6.7 & -1.33 $\pm$ 0.14 & -1.37 $\pm$ 0.13\\
NGC3073 & 0.97 $\pm$ 0.69 & FUV & 2.15 $\pm$ 0.37 & 35.8 & 8.9 & 8.9 & 7.2 $\pm$ \hspace{4pt}0.1 & 309 $\pm$ \hspace{4pt}1.3 & -1.49 $\pm$ 0.36 & -1.27 $\pm$ 0.35\\
NGC3156 & 4.23 $\pm$ 0.96 & PXIV & 1.04 $\pm$ 0.18 & 64.5 & 16.8 & 14.3 & 9.2 $\pm$ \hspace{4pt}0.3 & 147 $\pm$ \hspace{4pt}1.7 & -2.21 $\pm$ 0.15 & -2.37 $\pm$ 0.13\\
NGC3182 & 5.79 $\pm$ 1.76 & PXIV & 1.66 $\pm$ 0.21 & 39.2 & 11.8 & 33.1 & 27.4 $\pm$ \hspace{4pt}0.3 & 524 $\pm$ 21.2 & -1.62 $\pm$ 0.18 & -1.61 $\pm$ 0.17\\
NGC3245 & 0.09 $\pm$ 0.01 & HST & 2.30 $\pm$ 0.15 & 4.3 & 8.1 & 184.0 & 158.5 $\pm$ \hspace{4pt}1.2 & 173 $\pm$ \hspace{4pt}4.6 & 0.09 $\pm$ 0.10 & 0.29 $\pm$ 0.08\\
NGC3489 & 0.50 $\pm$ 0.18 & PXIV & 1.58 $\pm$ 0.22 & 18.3 & 12.3 & 108.0 & 67.7 $\pm$ \hspace{4pt}1.3 & 626 $\pm$ 13.7 & -1.19 $\pm$ 0.20 & -1.19 $\pm$ 0.19\\
NGC3599 & 0.05 $\pm$ 0.01 & HST & 2.71 $\pm$ 0.15 & 8.7 & 6.4 & 33.6 & 31.4 $\pm$ \hspace{4pt}0.3 & 9 $\pm$ \hspace{4pt}0.9 & 0.16 $\pm$ 0.11 & 0.36 $\pm$ 0.10\\
NGC3607 & 8.14 $\pm$ 1.36 & PXIV & 1.59 $\pm$ 0.17 & 24.2 & 12.3 & 105.9 & 38.5 $\pm$ \hspace{4pt}2.1 & 707 $\pm$ \hspace{4pt}5.4 & -2.00 $\pm$ 0.13 & -2.09 $\pm$ 0.11\\
NGC3619 & 1.60 $\pm$ 0.72 & PXIV & 2.36 $\pm$ 0.26 & 10.0 & 7.9 & 45.9 & 28.1 $\pm$ \hspace{4pt}0.7 & 687 $\pm$ \hspace{4pt}7.7 & -1.46 $\pm$ 0.24 & -1.37 $\pm$ 0.23\\
NGC3626 & 1.51 $\pm$ 0.08 & PXIV & 2.26 $\pm$ 0.15 & 35.8 & 8.4 & 166.7 & 156.4 $\pm$ \hspace{4pt}1.0 & -  & -0.71 $\pm$ 0.10 & - \\
NGC3665 & 8.84 $\pm$ 2.10 & PXIV & 2.16 $\pm$ 0.19 & 25.7 & 8.8 & 138.9 & 55.4 $\pm$ \hspace{4pt}1.2 & 181 $\pm$ 12.8 & -1.91 $\pm$ 0.15 & -2.15 $\pm$ 0.14\\
NGC4036 & 1.97 $\pm$ 0.73 & FUV & 1.84 $\pm$ 0.23 & 16.3 & 10.6 & 60.0 & 41.6 $\pm$ \hspace{4pt}0.8 & 256 $\pm$ \hspace{4pt}4.7 & -1.11 $\pm$ 0.20 & -1.13 $\pm$ 0.20\\
NGC4111 & 0.49 $\pm$ 0.22 & HST & 1.71 $\pm$ 0.26 & 17.1 & 11.4 & 96.7 & 84.5 $\pm$ \hspace{4pt}1.0 & 210 $\pm$ \hspace{4pt}3.7 & -0.59 $\pm$ 0.24 & -0.53 $\pm$ 0.23\\
NGC4119 & 1.55 $\pm$ 0.44 & PXIV & 1.95 $\pm$ 0.20 & 38.5 & 10.0 & 47.2 & 29.2 $\pm$ \hspace{4pt}1.2 & 82 $\pm$ 11.0 & -1.73 $\pm$ 0.17 & -1.93 $\pm$ 0.17\\
NGC4150 & 1.32 $\pm$ 0.33 & PXIV & 1.71 $\pm$ 0.19 & 41.7 & 11.5 & 72.7 & 67.1 $\pm$ \hspace{4pt}0.5 & 109 $\pm$ \hspace{4pt}2.4 & -1.29 $\pm$ 0.15 & -1.41 $\pm$ 0.14\\
NGC4203 & 0.32 $\pm$ 0.18 &  24$\mu$m & 2.04 $\pm$ 0.31 & 8.9 & 9.5 & 79.9 & 35.4 $\pm$ \hspace{4pt}1.2 & 546 $\pm$ 12.6 & -1.17 $\pm$ 0.29 & -1.09 $\pm$ 0.28\\
NGC4283 & 0.28 $\pm$ 0.17 & FUV & 1.66 $\pm$ 0.34 & 10.5 & 11.8 & 10.6 & 2.2 $\pm$ \hspace{4pt}0.3 & 72 $\pm$ \hspace{4pt}3.6 & -2.03 $\pm$ 0.32 & -2.01 $\pm$ 0.31\\
NGC4324 & 1.92 $\pm$ 0.06 & PXIV & 1.69 $\pm$ 0.15 & 79.8 & 11.6 & 46.9 & 30.5 $\pm$ \hspace{4pt}0.8 & 405 $\pm$ 22.9 & -1.76 $\pm$ 0.10 & -1.84 $\pm$ 0.08\\
NGC4429 & 0.98 $\pm$ 0.35 & PXIV & 2.40 $\pm$ 0.23 & 11.9 & 7.7 & 190.2 & 123.1 $\pm$ \hspace{4pt}5.1 & 464 $\pm$ 28.8 & -0.98 $\pm$ 0.20 & -0.99 $\pm$ 0.19\\
NGC4435 & 0.57 $\pm$ 0.27 & PXIV & 2.30 $\pm$ 0.27 & 8.1 & 8.2 & 111.9 & 69.5 $\pm$ \hspace{4pt}2.7 & 209 $\pm$ \hspace{4pt}4.7 & -0.98 $\pm$ 0.25 & -1.01 $\pm$ 0.24\\
NGC4459 & 1.91 $\pm$ 0.48 & PXIV & 1.96 $\pm$ 0.19 & 11.4 & 9.9 & 142.3 & 97.0 $\pm$ \hspace{4pt}2.2 & 418 $\pm$ \hspace{4pt}6.9 & -1.29 $\pm$ 0.15 & -1.36 $\pm$ 0.14\\
NGC4476 & 2.65 $\pm$ 0.61 & PXIV & 1.63 $\pm$ 0.18 & 36.5 & 12.0 & 30.3 & 23.3 $\pm$ \hspace{4pt}0.2 & 149 $\pm$ \hspace{4pt}4.5 & -1.93 $\pm$ 0.15 & -2.13 $\pm$ 0.14\\
NGC4477 & 0.28 $\pm$ 0.19 & PXIV & 2.10 $\pm$ 0.36 & 5.6 & 9.2 & 44.6 & 10.6 $\pm$ \hspace{4pt}1.5 & 408 $\pm$ \hspace{4pt}7.2 & -1.35 $\pm$ 0.34 & -1.13 $\pm$ 0.33\\
NGC4526 & 2.22 $\pm$ 0.52 & PXIV & 2.24 $\pm$ 0.18 & 14.9 & 8.4 & 349.9 & 261.4 $\pm$ \hspace{4pt}8.4 & 658 $\pm$ \hspace{4pt}7.0 & -1.00 $\pm$ 0.15 & -1.04 $\pm$ 0.14\\
NGC4596 & 1.10 $\pm$ 0.37 &  24$\mu$m & 1.27 $\pm$ 0.22 & 0.8 & 14.7 & 39.8 & 4.2 $\pm$ \hspace{4pt}1.9 & 330 $\pm$ \hspace{4pt}6.5 & -2.21 $\pm$ 0.25 & -1.98 $\pm$ 0.18\\
NGC4643 & 0.85 $\pm$ 0.33 & FUV & 1.34 $\pm$ 0.24 & 0.7 & 14.1 & 87.6 & 48.9 $\pm$ \hspace{4pt}1.6 & 31 $\pm$ \hspace{4pt}8.9 & -1.25 $\pm$ 0.21 & -1.37 $\pm$ 0.24\\
NGC4684 & 0.38 $\pm$ 0.17 & FUV & 1.63 $\pm$ 0.26 & 22.8 & 12.0 & 247.7 & 238.0 $\pm$ \hspace{4pt}1.0 & 977 $\pm$ \hspace{4pt}9.5 & -0.41 $\pm$ 0.24 & -0.29 $\pm$ 0.23\\
NGC4694 & 1.00 $\pm$ 0.35 & PXIV & 2.06 $\pm$ 0.22 & 73.0 & 9.3 & 112.0 & 97.9 $\pm$ \hspace{4pt}0.8 & 778 $\pm$ \hspace{4pt}8.7 & -1.23 $\pm$ 0.20 & -1.24 $\pm$ 0.19\\
NGC4710 & 2.97 $\pm$ 0.07 & PXIV & 2.61 $\pm$ 0.15 & 91.5 & 6.8 & 416.1 & 383.9 $\pm$ \hspace{4pt}4.6 & 108 $\pm$ 15.7 & -0.96 $\pm$ 0.10 & -1.03 $\pm$ 0.10\\
NGC4753 & 4.76 $\pm$ 1.07 & PXIV & 2.02 $\pm$ 0.18 & 33.4 & 9.6 & 250.5 & 110.3 $\pm$ \hspace{4pt}4.3 & 100 $\pm$ \hspace{4pt}7.6 & -1.57 $\pm$ 0.15 & -1.76 $\pm$ 0.14\\
NGC5173 & 4.58 $\pm$ 1.76 & FUV & 2.04 $\pm$ 0.24 & 41.1 & 9.5 & 18.1 & 8.5 $\pm$ \hspace{4pt}0.2 & 518 $\pm$ \hspace{4pt}6.9 & -2.23 $\pm$ 0.21 & -2.09 $\pm$ 0.20\\
NGC5273 & 0.85 $\pm$ 0.32 &  24$\mu$m & 1.38 $\pm$ 0.23 & 21.7 & 13.8 & 83.9 & 81.7 $\pm$ \hspace{4pt}0.5 & 161 $\pm$ \hspace{4pt}3.5 & -0.48 $\pm$ 0.21 & -0.40 $\pm$ 0.20\\
NGC5379 & 4.85 $\pm$ 1.43 & FUV & 1.84 $\pm$ 0.20 & 131.2 & 10.6 & 39.3 & 30.0 $\pm$ \hspace{4pt}0.3 & 245 $\pm$ 13.9 & -2.05 $\pm$ 0.17 & -2.26 $\pm$ 0.16\\
NGC5866 & 2.39 $\pm$ 0.05 & PXIV & 2.38 $\pm$ 0.15 & 58.7 & 7.8 & 225.7 & 210.5 $\pm$ \hspace{4pt}1.4 & 495 $\pm$ \hspace{4pt}7.8 & -0.67 $\pm$ 0.10 & -0.64 $\pm$ 0.08\\
NGC6014 & 3.71 $\pm$ 1.53 & PXIV & 2.20 $\pm$ 0.25 & 19.1 & 8.6 & 130.0 & 105.1 $\pm$ \hspace{4pt}0.7 & 430 $\pm$ 18.7 & -1.57 $\pm$ 0.22 & -1.72 $\pm$ 0.21\\
NGC6798 & 0.69 $\pm$ 0.55 & FUV & 2.45 $\pm$ 0.41 & 13.4 & 7.5 & 14.5 & 8.2 $\pm$ \hspace{4pt}0.3 & 17 $\pm$ \hspace{4pt}6.6 & -1.23 $\pm$ 0.40 & -1.33 $\pm$ 0.43\\
NGC7465 & 10.42 $\pm$ 2.04 & PXIV & 2.15 $\pm$ 0.17 & 64.0 & 8.9 & 313.3 & 310.2 $\pm$ \hspace{4pt}1.8 & -  & -0.88 $\pm$ 0.14 & - \\
PGC016060 & 1.22 $\pm$ 0.04 &H$\beta$ & 2.18 $\pm$ 0.15 & 25.3 & 8.8 & 23.2 & 18.8 $\pm$ \hspace{4pt}0.2 & -  & -1.31 $\pm$ 0.10 & - \\
PGC029321 & 3.91 $\pm$ 1.73 & PXIV & 2.02 $\pm$ 0.26 & 45.1 & 9.6 & 342.6 & 341.7 $\pm$ \hspace{4pt}2.5 & -  & -0.47 $\pm$ 0.23 & - \\
PGC056772 & 0.97 $\pm$ 0.83 & FUV & 2.21 $\pm$ 0.44 & 28.2 & 8.6 & 124.2 & 122.7 $\pm$ \hspace{4pt}0.6 & 37 $\pm$ \hspace{4pt}6.2 & -0.36 $\pm$ 0.43 & -0.28 $\pm$ 0.43\\
PGC058114 & 1.12 $\pm$ 0.55 & PXIV & 2.70 $\pm$ 0.28 & 1.8 & 6.5 & 504.8 & 502.3 $\pm$ \hspace{4pt}3.0 & 115 $\pm$ 13.2 & -0.26 $\pm$ 0.26 & -0.16 $\pm$ 0.25\\
PGC061468 & 5.19 $\pm$ 1.75 & H$\beta$ & 1.28 $\pm$ 0.22 & 57.4 & 14.6 & 15.3 & 14.2 $\pm$ \hspace{4pt}0.2 & -  & -1.78 $\pm$ 0.19 & - \\
UGC05408 & 1.26 $\pm$ 0.43 & PXIV & 2.53 $\pm$ 0.22 & 34.9 & 7.2 & 218.0 & 216.9 $\pm$ \hspace{4pt}1.0 & 878 $\pm$ \hspace{4pt}9.7 & -0.17 $\pm$ 0.19 & 0.01 $\pm$ 0.18\\
UGC06176 & 1.87 $\pm$ 0.47 & PXIV & 2.64 $\pm$ 0.19 & 10.2 & 6.7 & 232.7 & 230.3 $\pm$ \hspace{4pt}1.4 & -  & -0.36 $\pm$ 0.15 & - \\
UGC09519 & 2.52 $\pm$ 0.98 & PXIV & 2.41 $\pm$ 0.24 & 25.0 & 7.7 & 30.4 & 27.5 $\pm$ \hspace{4pt}0.2 & -  & -1.45 $\pm$ 0.21 & - \\
\hline 
\end{tabular*}
\parbox[t]{\textwidth}{ \textit{Notes:} Column one lists the name of the galaxy. Column 2 contains the area of the star-forming region, estimated from the source listed in Column 3. PXIV refers to Table 1 of \cite{2013MNRAS.429..534D}, A11 refers to \cite{2011ApJ...735...88A}, H$\beta$ refers to a size calculated from the Balmer line emitting region visible in SAURON observations. Column 5 lists the total gas surface density derived from the H$_2$ and \hi\ masses of these objects, as described in the text. Column 5 lists the dynamical time at the outer edge of the molecular disc, calculated from the circular velocity of these galaxies at this radius (see \citealt{2011MNRAS.414..968D} and Paper XIV). Column 6 contains the local free-fall time of the gas, calculated as in Equation \ref{tff_eqn}. Column 7 contains the observed \textit{WISE} integrated 22~\mum\ flux density of the object, before correction for circumstellar emission. Column 8 contains the \textit{WISE} integrated 22~\mum\ flux density corrected for circumstellar emission using Equations \ref{my_passivecorr} and \ref{my_passivecorr2}. Column 9 contains the integrated FUV flux density of the object, after correction for Galactic extinction \citep{1998ApJ...500..525S}. A dash in this column indicates that no measurements are available. Column 10 contains the logarithm of the SFR surface density estimated using the equation in \cite{2007ApJ...666..870C}, after correction for circumstellar emission.  Column 11 contains the logarithm of the SFR surface density estimated from corrected 22~\mum\ fluxes and \textit{GALEX} FUV photometry, using the relation from \cite{2011ApJ...741..124H}. }
\label{obstable}
\end{table*}

\subsubsection{Literature Comparison}

As part of the \atlas\ survey, we have also estimated SFRs in some of these objects from \textit{Spitzer} observations of (non-stellar) 8 \mum\ emission (Falc\'on-Barroso et al, in prep; including the earlier results of \citealt{2010MNRAS.402.2140S}).

Twenty-three of our molecular gas-rich sample have \textit{Spitzer} measurements. The scatter between the 8~\mum\ and 22~\mum\ measures of star formation is larger than that between the two 22 \mum-based measures discussed above, but generally the agreement is good, with a scatter of $\approx$0.4 dex. The 8 \mum\ SFRs were estimated using the calibration of \cite{2005ApJ...632L..79W}, that is based upon a different galaxy sample, and this may be the cause of the larger scatter.

 \subsection{Kennicutt-Schmidt relations}

   \begin{figure*}
\begin{center}
\includegraphics[width=0.75\textwidth,angle=0,clip,trim=0.0cm 0.0cm 0cm 0cm]{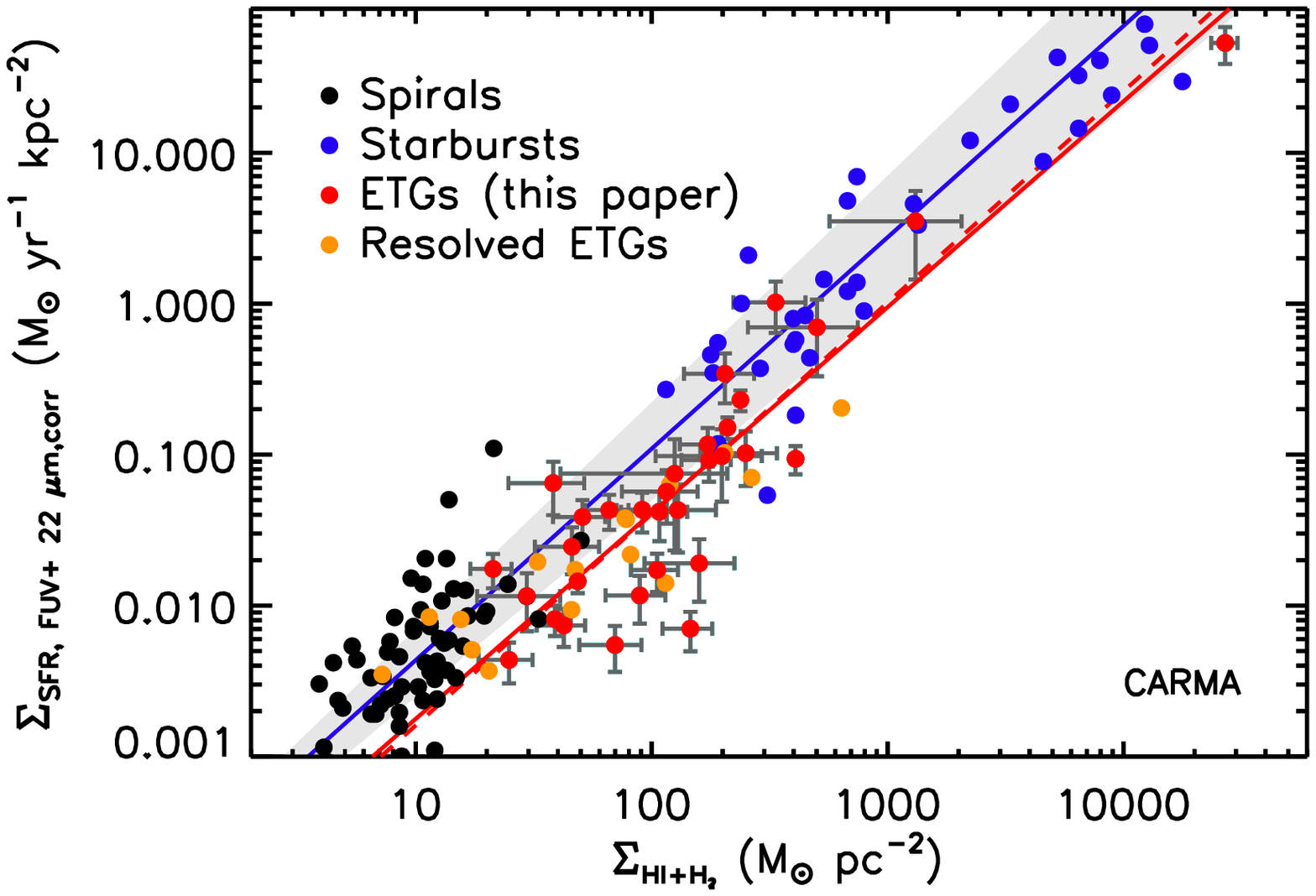}
 \end{center}
 \caption{As the bottom-right panel of Figure \ref{WISE22umfig}, but showing for reference the spiral and starburst objects of K98 (where the SFRs were calculated from H$\alpha$ emission, and have been corrected to a Kroupa IMF) and the spatially-resolved star formation rates (in radial bins) of 6 ETGs (presented in Paper XXII, where the SFR was calculated from non-stellar 8\mum\ emission).  }
 \label{ksfig_allgals}
 \end{figure*}

 Star-formation relations are usually given with respect to star-formation rate and gas surface densities, i.e. $\Sigma_{\rm SFR} \propto  \Sigma_{\rm HI+H2}^n$, where $n$ is some power-law exponent. This is physically motivated by theoretical predictions that star formation depends on gas volume density \citep[e.g.][]{1959ApJ...129..243S}, and additionally removes the distance dependence of the relation. For a sample of local star-forming spiral and starburst galaxies, \cite{1998ApJ...498..541K} (hereafter K98) found $n$ = 1.4 (the so-called Kennicutt-Schmidt relation), as shown below after correction to a \cite{2001MNRAS.322..231K} IMF:
 
 \begin{eqnarray}
\log_{10}\left(\Sigma_{\rm SFR}\right)=(1.4\pm0.15) \log_{10}\left(\Sigma_{\rm H{\small I}+H_2}\right)  - (3.76\pm0.12),
\label{k98law}
\end{eqnarray}

\noindent where $\Sigma_{\rm SFR}$ is in units of $M_{\odot}$ yr$^{-1}$ kpc$^{-2}$ and $\Sigma_{\rm HI+H_2}$ is in units of $M_{\odot}$ pc$^{-2}$.
 A value of $n$ greater than unity implies that the star-formation efficiency (SFE; where SFE $\equiv$ SFR/$M_{\rm gas}$) increases in high column density clouds. 
 Other works studying star formation within spatially-resolved regions in nearby spiral galaxies suggest a constant SFE (i.e. $n\approx$ 1; \citealt{1991ARA&A..29..581Y,1996AJ....112.1903Y,2008AJ....136.2846B}). We compare our ETGs to both the Kennicutt-Schmidt relation and the constant SFE relation of \citet[][hereafter B08]{2008AJ....136.2846B} in Figures \ref{WISE22umfig} and \ref{WISE22umfig_bigiel}.
 
In addition to the controversy surrounding the slope of the Kennicutt-Schmidt relation, it seems that high-redshift starburst galaxies form more stars per unit gas mass than their local analogues  \citep[e.g.][hereafter D10+G10]{2010ApJ...714L.118D,2010MNRAS.407.2091G}. This has led to the suggestion that two different SF regimes exist: a long-lasting mode for discs (evolving secularly) and a more rapid mode for rapidly star-forming high-redshift objects (with major mergers and/or dense SF regions). We investigate where our sample of ETGs falls with respect to these two SF modes in this Section.

 Figure \ref{WISE22umfig} shows the surface density of star formation (derived using 22~\mum\ emission only in the top row of panels, and the combined FUV+22~\mum\ calibration in the bottom row), plotted against the surface density of H$_2$ and \hi\ in our H$_2$-rich galaxy sample (calculated as described in Section \ref{data}).
 We do not show plots with the molecular gas only, as these objects are molecule-dominated and hence the derived star-formation relations are almost identical.  
  The plots in the left column show all our sample of H$_2$-rich ETGs, while the plots in the right column show only those objects where CO interferometry is available, leading to a better determination of the total CO flux and directly-measured molecular gas reservoir sizes.  Figure \ref{WISE22umfig} also shows the star-formation relation of K98 as a blue line, with the 1$\sigma$ error region shaded in grey. 
 The best fit to our data points is shown as a solid red line, while the best fit with a slope fixed to that found by K98 is shown as a dashed red line. The coefficients of these fits are shown in the figure legends, and are reproduced in Table \ref{ksfittable}. Figure \ref{WISE22umfig_bigiel} is analogous to Figure \ref{WISE22umfig}, with the constant SFE relation of B08 shown for comparison (rather than K98).

  Figure \ref{ksfig_allgals} shows our ETGs and the spiral and starburst galaxies of K98 on the same plot for comparison. We show only the galaxies from this work with CO interferometry available, and use SFRs derived from the combination of \textit{WISE} and \textit{GALEX} daya.  We also include the ETGs from Paper XXII, where we have spatially-resolved star formation rates (in radial bins) for 6 of our sample ETGs, calculated from non-stellar 8 \mum\ emission. Our trend based on global measurements agrees well with the resolved observations presented in that paper (although the best-fit slope to the PXXII sample would be slightly shallower). 
 
 Although generally within the scatter of the original K98 relation, it is clear for all indicators that our ETGs have a lower average SFE than both the spiral and starburst galaxies making up the sample of K98 (and thus a much lower SFE than the high-redshift objects of D10+G10). The left column of plots in Figure \ref{WISE22umfig} (which includes all objects) shows increased scatter, as expected given the larger uncertainties on the reservoir areas and masses, but they still suggest that the SFE of ETGs is lower than that of later-type objects. The zero points of the best-fit relations with a fixed slope (listed in Table \ref{ksfittable}) suggest a relation offset by a factor of between 2.2 and 2.5 from that of K98 (depending on the tracer/sample selection), and a factor $\approx$17 from that of the high-redshift starbursts. These mean offsets are significant at greater than a $3\sigma$ level, even given the large scatter in the observations. Looking at the galaxies individually, it is clear that this effect is dominated by a specific set of objects, whose properties will be discussed further below.

 The slopes of our best-fit relations when using only 22~\mum\ as a tracer of star formation are slightly shallower than the relation of K98, with slopes of $n$=1.19$\pm$0.03 and 1.11$\pm$0.04 (when fitting all galaxies and those with interferometric data only, respectively). These are still steeper than a constant SFE relation, as can clearly be seen in Figure \ref{WISE22umfig_bigiel}. When using a calibration with both FUV and 22~\mum\ fluxes our best fits are steeper, with $n$=1.49$\pm$0.04 and 1.31$\pm$0.04, respectively, consistent with the slope found by K98.
  B08 suggest that when one investigates star formation in a spatially-resolved fashion (rather than in an integrated manner as done here), one obtains a shallower relation. The slope obtained with resolved observations in Paper XXII is indeed shallower (as seen in Figure \ref{ksfig_allgals}). However, this result is still the subject of some debate \citep[e.g.][]{2013ApJ...772L..13M}, and we will investigate this matter further when presenting spatially-resolved star-formation relations for all these galaxies in a future work.

\begin{table}
\caption{Kennicutt-Schmidt relation fits}
\begin{tabular*}{0.5\textwidth}{@{\extracolsep{\fill}}l r r r r}
\hline
SF Indicator  & Sample & $n$ &  $c$ & $\chi^2_{\rm red}$\\
  &  & &   $\log$($M_{\odot}$ yr$^{-1}$ kpc$^{-2}$) &   \\
 (1) & (2) & (3) & (4) & (5) \\
 \hline
 22$\mu$m & all & $  1.20\pm 0.04$&$   -3.58\pm 0.09$     &  6.74\\
 22$\mu$m & all & $  1.40$&$   -4.01\pm 0.02$     &  7.12\\
 22$\mu$m & PXVIII & $  1.14\pm 0.04$&$   -3.57\pm 0.10$     &  3.83\\
 22$\mu$m & PXVIII & $  1.40$&$   -4.15\pm 0.03$     &  4.66\\
FUV + 22$\mu$m & all & $  1.51\pm 0.04$&$   -4.20\pm 0.10$     &  9.90\\
FUV + 22$\mu$m & all & $  1.40$&$   -3.96\pm 0.02$     & 10.02\\
FUV + 22$\mu$m & PXVIII & $  1.36\pm 0.05$&$   -4.11\pm 0.11$     &  4.23\\
FUV + 22$\mu$m & PXVIII & $  1.40$&$   -4.19\pm 0.03$     &  4.25\\
 \hline
 \end{tabular*}
\parbox[t]{0.5 \textwidth}{ \textit{Notes:} This table contains the fitted slope and intercept for the Kennicutt-Schmidt relations presented in Figure \ref{WISE22umfig} (here paramaterised as {log$_{10}$($\Sigma_{\rm SFR}$)=$n$~log$_{10}$($\Sigma_{\rm gas}$)~+~$c$}). Column 1 lists the SFR indicator used, and Column 2 the sample of galaxies included in the fit. PXVIII refers to the interferometrically mapped sample of \cite{2013MNRAS.432.1796A}. The slope ($n$) and intercept ($c$) of the best fits are given in Columns 3 and 4. Where the slope was fixed to $n$~=~1.4 (the best fit value of K98), this is indicated in Column 3 and no error bar is reported on the slope. Column 5 shows the reduced $\chi^2$ for each fit, indicating how well the best fit values represent the observed data points. }
\label{ksfittable}
\end{table}

\subsection{Elmegreen-Silk relation}
\label{eslaw}
An alternative parameterisation of the relation between SFR and gas surface density depends on the dynamical time at the edge of the star-forming gas disc ($\tau_{dyn}$), as shown below \citep{1997RMxAC...6..165E,1997ApJ...481..703S}: 

 \begin{equation}
\Sigma_{\rm SFR} \propto  (\Sigma_{\rm HI+H2/\tau_{dyn}})^n.
 \end{equation}
 
\noindent K98 also placed their sample of spiral and starburst galaxies on this relation (estimating the radial extent of the gas by finding the edge of the main H$\alpha$ or Br$\gamma$-emitting disc). They found a tight linear correlation, that is an equally good description of the data points as Equation \ref{k98law}. Their best-fit relation is (again after correction to a Kroupa IMF): 

 \begin{equation}
\frac{\Sigma_{\rm SFR}}{M_{\odot}~\rm yr^{-1}~pc^{-2}}=0.106 \left(\frac{\Sigma_{\rm H{\small I}+H_2}}{M_{\odot}~\rm pc^{-2}}\right)\left(\frac{\tau_{\rm dyn}}{\rm yr}\right)^{-1},
\label{k98law_tdyn}
\end{equation}

\noindent We estimate the dynamical times of our sample galaxies using the \atlas\ mass models from Paper XV, from which we can extract a circular velocity ($V_{\rm circ}$) profile as a function of radius, assuming a mass-follows-light model (models A of Paper XV). Although our galaxies contain dark matter, as well as stellar matter, this contributes only 13\% in median within R$_{\rm e}$, which is generally larger than the region where we detect CO. This implies that the total mass profile has a slope very close to that of the stellar distribution alone, justifying our use of mass-follows-light models. We define the dynamical time at the outer edge of the gas disk (as determined in Paper XIV) simply as $\tau_{\rm dyn}$~=~$2\pi$R/V$_{\rm circ}$. The dynamical times for our objects are listed in Table \ref{obstable}. Paper XIV has shown that the molecular gas is dynamically cold and follows well the circular velocity profile in the majority of our objects, and hence this $\tau_{dyn}$ measurement should provide good estimates of the dynamical times within the molecular gas itself.

\begin{figure}
\begin{center}
\includegraphics[width=0.45\textwidth,angle=0,clip,trim=0.0cm 0cm 0cm 0cm]{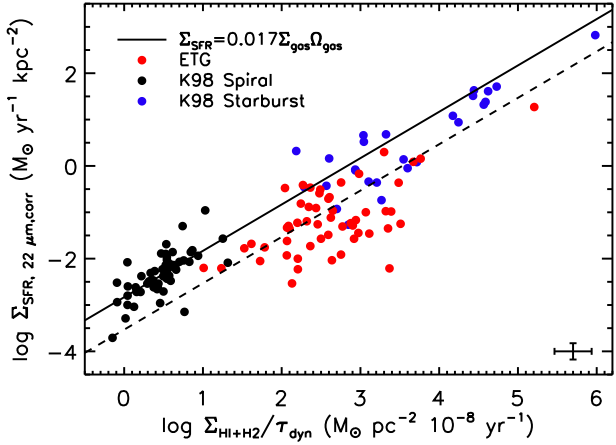}
 \end{center}
 \caption{As the top-right panel of Figure 2, but with the gas surface densities divided by the dynamical time (estimated at the outer edge of the molecular gas disc). Only galaxies with interferometic CO maps available are included. Molecular gas-rich \atlas\ ETGs are the red circles, while the spiral and starburst objects of K98 are shown in black and red, respectively. The best fit reported in K98 is shown as a black line, and the best fit to our points, assuming the same slope as K98, is shown as a dashed line. The intercept of the best fit line corresponds to ETGs turning {$\approx$2\%} of their gas into stars per dynamical time, a factor of $\approx$5 lower than spiral/starburst galaxies. The mean error bar on the ETG data points is shown in the bottom right corner.}

 \label{tdynplot}
 \end{figure}

Figure \ref{tdynplot} shows the position of our molecular gas-rich sample ETGs (red circles) with respect to the Elmegreen-Silk (E-S) relation of K98 (as in Equation \ref{k98law_tdyn}). Also plotted for comparison are the spiral (black circles) and starburst (blue circles) sample of K98.

Our ETGs fall systematically below the E-S relation, with a large scatter. 
The best fit to our sample (assuming the same linear slope as K98) is $\Sigma_{\rm SFR}=2.96\times10^{-3}\Sigma_{\rm gas}\Omega_{\rm gas}$, suggesting ETGs turn $\approx$2\% of their gas into stars per dynamical time, a factor of $\approx$6 lower than spiral/starburst galaxies (and high-redshift starbursts which are also found to obey the E-S relation; e.g. D10+G10). 
The ETGs fall in the gap between the spiral galaxies and starburst nuclei on this plot, in the same region as spiral galaxy centres (as shown in K98), but they are offset to lower SFRs. 
The cause of this effect will be discussed further in Section \ref{dyn_sf_discuss}.

\subsection{Local volumetric star-formation relations}
\label{kdmsfr_rel}
An alternative approach to unifying star-formation relations is to assume that star formation efficiency is set by the local value of the gas free-fall time. 
As discussed in depth in \citet[][hereafter KDM12]{2012ApJ...745...69K}, such a volumetric star-formation relation can reproduce observations of both Galactic and extragalactic star forming sources. The local free-fall time that is relevant for star formation can be calculated in several ways, depending on whether the ISM of the object is assumed to be in small bound units (such as giant molecular clouds; GMCs) or a continuous sheet with local density variations. The former is thought to be a more physical model of low-redshift galaxies, while the later is though to be appropriate in high-redshift and starbursting sources. 

The fundamental parameters that vary in the GMC based free-fall time ($t_{\rm ff,GMC}$) estimate are the gas velocity dispersion and the observed gas density (see Equation \ref{tff_eqn} below). 
No study of the molecular gas velocity dispersion in ETGs currently exist, but Davis et al., in prep., suggest that this dispersion is low, and likely similar to local spirals which have $\sigma_{\rm gas}\ltsimeq$~12~\kms \citep{2013arXiv1309.6324C}.  
Assuming this velocity dispersion does not strongly vary between sources, the GMC based estimate of free-fall time just depends on the gas surface density itself (making this correction factor a simple rotation of the points in log-space).

\begin{equation}
\label{tff_eqn}
t_{\rm ff,GMC}=\frac{\pi^{\frac{1}{4}}}{\sqrt{8}} \frac{\sigma_{\rm gas}}{G(\overline{\Sigma_{\rm GMC}}^3 \Sigma_{\rm gas})^{\frac{1}{4}}}
\end{equation}
\noindent where G is the gravitational constant, $\sigma_{gas}$ is the gas velocity dispersion (assumed here to be a constant 8 \kms\, as in KDM12), $\Sigma_{gas}$ is the observed (galaxy average) gas surface density and $\overline{\Sigma_{\rm GMC}}$ is the average GMC surface density, which we here assume is a constant 85 $M_{\odot}$ pc$^{-2}$, as in \cite{2012ApJ...745...69K}.

The alternative (starburst) prescription from KDM12 assumes that star formation is regulated by the dynamical stability of a continuous disc of gas, that globally should have a Toomre $Q$ parameter \citep{1964ApJ...139.1217T} of $\approx$1. In this case the parameters needed to calculate the free-fall time are the dynamical time (which enters the equation linearly), and the logarithmic derivative of the rotation curve ($\beta$=$\frac{\delta \mathrm{ln(V)}}{\delta \mathrm{ln(R)}}$) which enters to the power of -0.5. %
In objects where the majority of the gas reaches beyond the turnover of the galactic rotation curve $\beta \approx$0, and the free fall time simply depends on the dynamical time (as in the E-S star-formation relation discussed in Section \ref{eslaw}). 

We calculate the local free-fall times for our ETGs using the relations of KDM12.  We find that the GMC estimate (the functional form of which is shown in Equation \ref{tff_eqn}) is shorter in all objects, and hence dominant. This is expected, as local ETGs are usually not starbursts, and have been observed to have most of their molecular mass in discreet GMCs \citep{2013Natur.494..328D}. The adopted free-fall times are listed in Column 5 of Table \ref{obstable}.

\begin{figure}
\begin{center}
\includegraphics[width=0.45\textwidth,angle=0,clip,trim=0.0cm 0cm 0cm 0cm]{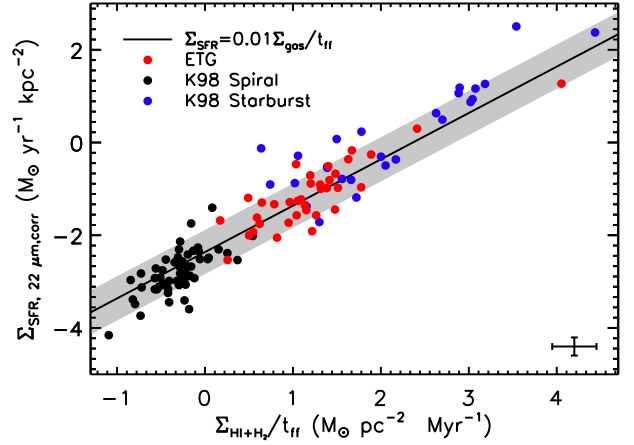}
 \end{center}
 \caption{As the top-right panel of Figure 2, but with the gas surface densities divided by the local free-fall time (estimated using Equation \ref{tff_eqn}). Only galaxies with interferometic CO maps available are included. Our \atlas\ ETGs are the red circles, while the spiral and starburst objects of K98 (with free-fall times as calculated in KDM12) are shown in black and red, respectively. The best fit reported in KDM12 is shown as a solid black line. It provides a good fit to our ETGs, so we do not plot our own fitted relation. The mean error bar on the ETG data points is shown in the bottom right corner.}
 \label{tffplot}
 \end{figure}

Using the free-fall time calculated from equation \ref{tff_eqn}, in Figure \ref{tffplot} we plot our ETGs on the local volumetric star-formation relation of KDM12. Also plotted for comparison are the spiral (black circles) and starburst sample (blue circles) from K98 (with free-fall times as calculated by \citealt{2012ApJ...745...69K}). Our ETGs fall onto the relation of KDM12, suggesting that on average they convert $\approx$1\% of their gas into stars per \textit{local free-fall time}. We discuss this result further in Section \ref{dyn_sf_discuss}.

\subsection{Dynamical drivers of star formation suppression}
\label{tdep_turnover}

 \begin{figure}
\begin{center}
\subfigure{\includegraphics[height=6cm,angle=0,clip,trim=0.0cm 0cm 0cm 0cm]{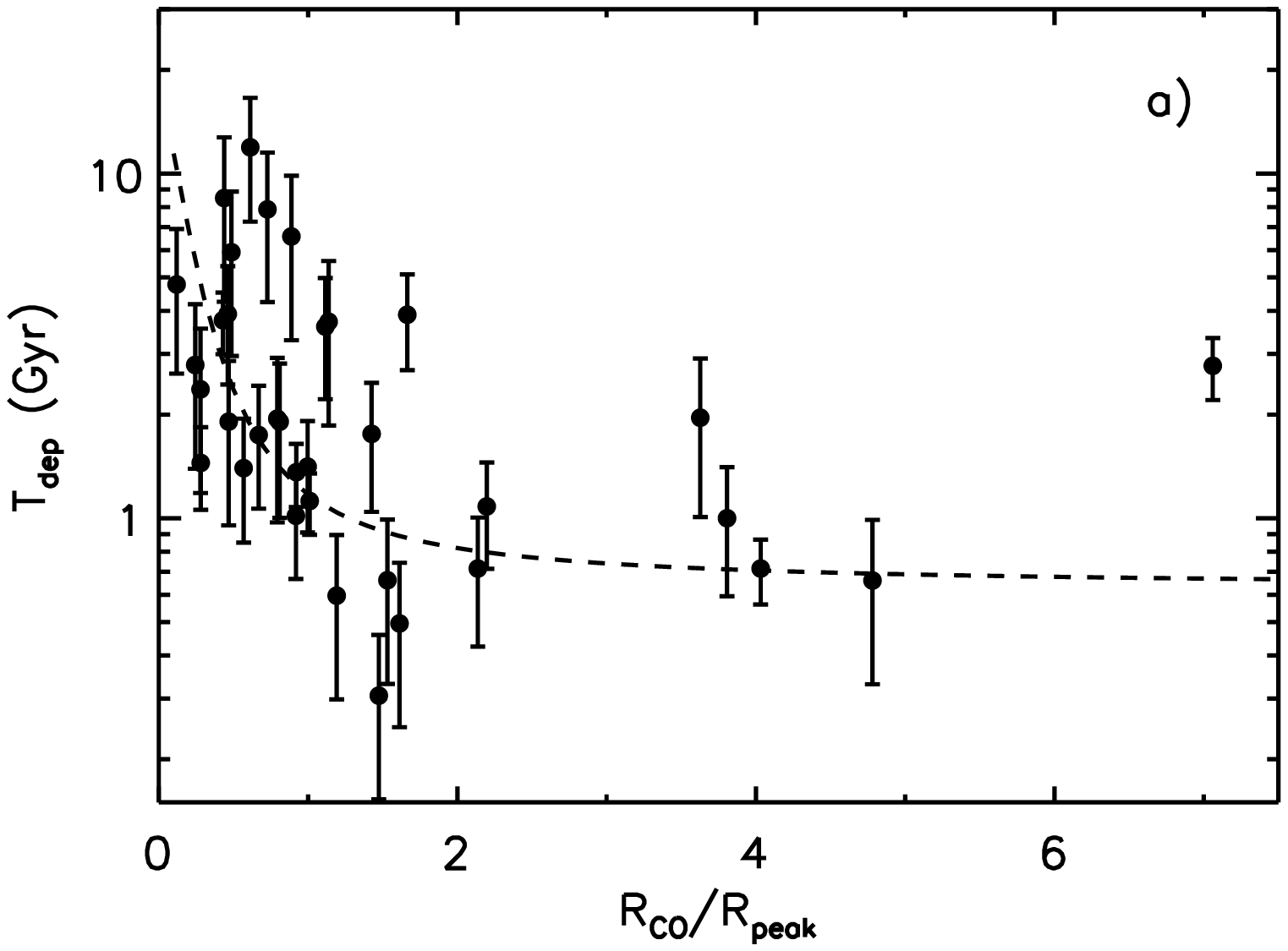}}
\subfigure{\includegraphics[height=6cm,angle=0,clip,trim=0.0cm 0cm 0cm 0cm]{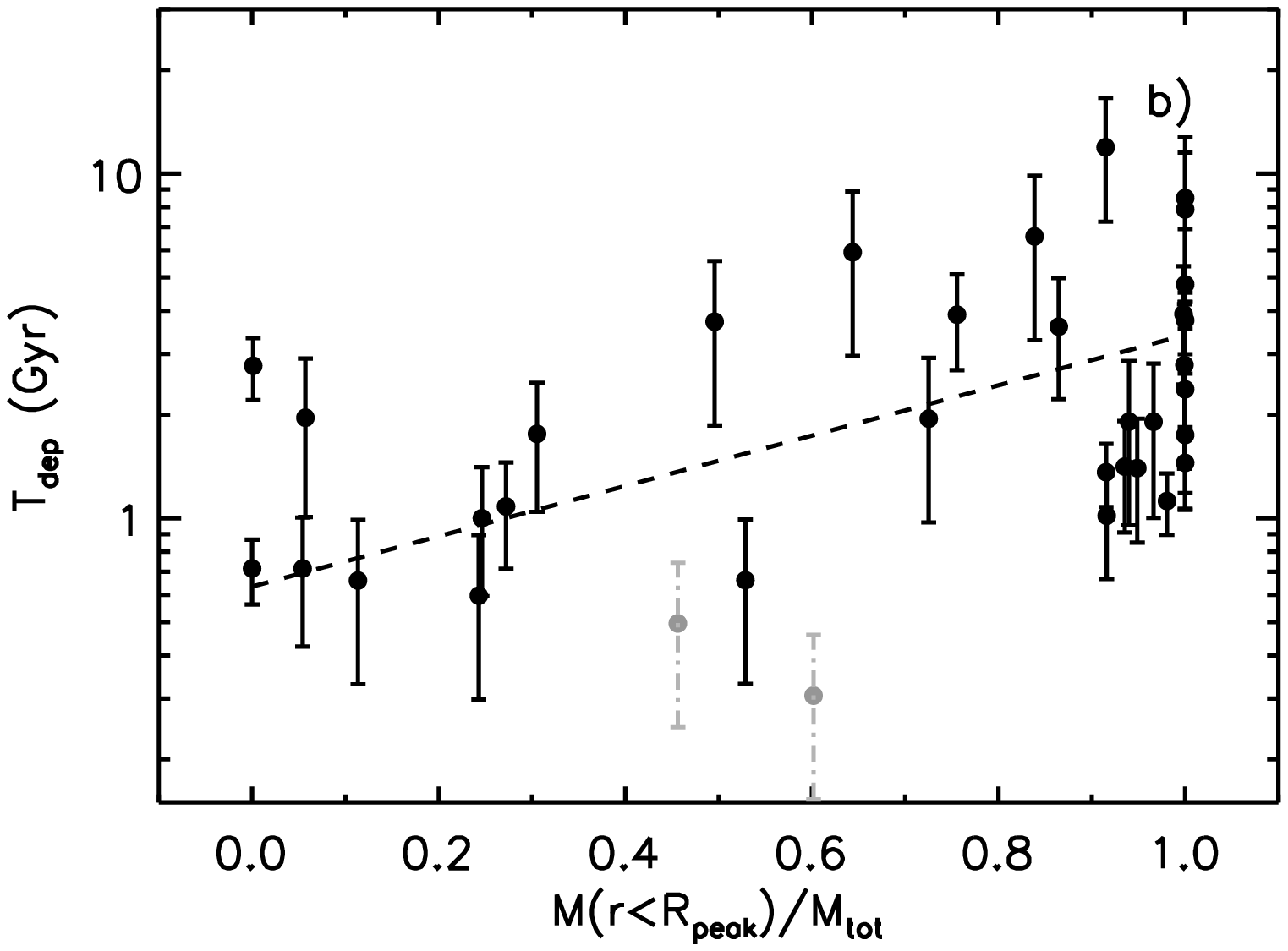}}
\subfigure{\includegraphics[height=6cm,angle=0,clip,trim=0.0cm 0cm 0cm 0cm]{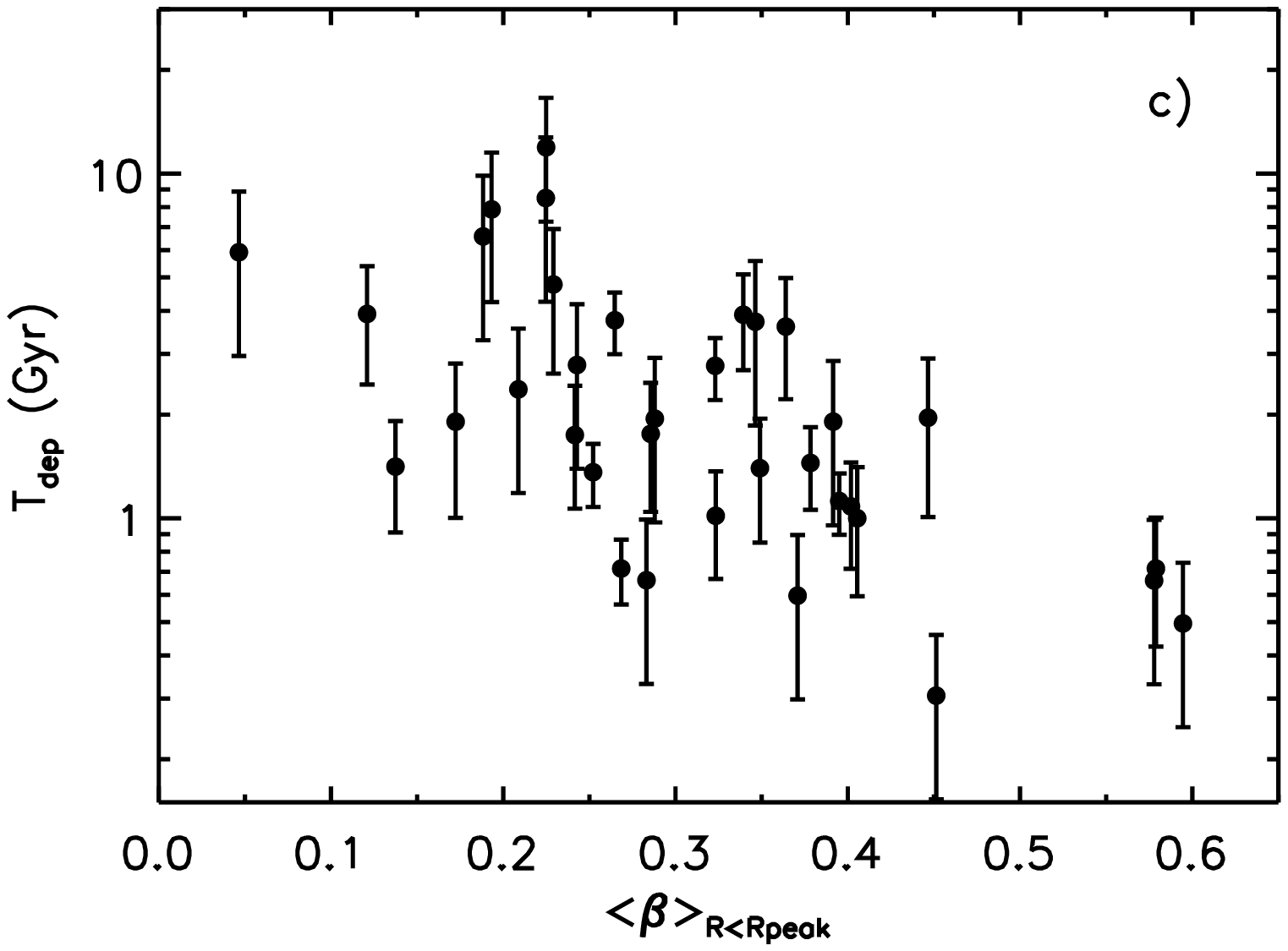}}
 \end{center}
 \caption{Gas depletion times plotted in panel a) as a function of the extent of the molecular gas, normalised by the location of the turnover in the galaxy rotation curve. The dashed line shows a simple mixing model as described in the text, where gas depletion times are assumed to be a Hubble time in the rising part of the rotation curve, and $\approx$0.8 Gyr in the flat part. Panel b) shows the fraction of the molecular gas mass present inside the turnover radius. The grey points are galaxies where within our beam size we cannot determine this quantity accurately. The dashed line is a guide to eye fitted to the solid points. Panel c) shows the mean logarithmic derivative of the rotation curve in the inner part of the rotation curve (inside R$_{\rm peak}$). }
 \label{tdeprotcurve}
 \end{figure}

The fact that local ETGs do not follow the same relationship between star formation and gas surface density in Figure \ref{WISE22umfig} suggests that there is some difference in the way star formation proceeds in these objects. Figure \ref{tffplot} suggests this may be a dynamical effect, as only when accounting for the relevant local timescale is a universal SF relation obtained, where molecule-rich ETGs form the same fraction of stars per free-fall time as nearby spiral and starbursting objects.  The obvious difference between ETGs and local spiral galaxies is that they tend to have, deeper, steeper potential wells. In this section we investigate if the shape of the potential correlates with the suppression of star formation discussed above. 

In Figure \ref{tdeprotcurve} (panel a) we plot the gas depletion time ($\equiv$M$_{\rm gas}$/SFR = 1/SFE) for those sample galaxies which were mapped in Paper XVIII, against the extent of the molecular gas (tabulated in Paper XIV) normalised by the location of the turnover in the rotation curve of that object (estimated from the JAM model circular velocity profiles published in Paper XV, as in \citealt{2011MNRAS.414..968D} and Paper XIV). Figure \ref{tdeprotcurve} shows that galaxies with long depletion times (and thus low SFE) have the majority of their molecular gas confined within regions where the rotation curve of the galaxy is still rising. The rest of the objects, which have more of their gas within the flat part of the galaxy rotation curve, have depletion times and star formation efficiencies consistent with those found for normal spiral galaxies ($\approx$0.5-1.5 Gyr). 

We also plot in Figure \ref{tdeprotcurve}, panel a) a simple mixing model (shown as a dashed line). This toy model assumes that two star formation regimes exist within early-type galaxies, one with very low SFE (which is valid in the rising part of the rotation curve), and one with a normal SFE, similar to that found in local spirals (which is valid in the flat part of the galaxy rotation curve). We assume depletion times of 0.8 Gyr for the normal regime, and a Hubble time for the low SFE regime respectively. We then assume all the gas is distributed in an exponential disc (which is only true for $\approx$50\% of these sources; Paper XIV), and that the maximum gas extent we can measure corresponds to 3 scale lengths. We then vary the scale length of this disc with respect to the turnover of rotation curve, and calculate a model "integrated" depletion time by weighting the two assumed depletion times by the fraction of gas in each regime. This leads to the curve shown in \ref{tdeprotcurve}. This toy model is likely to be a vast over-simplification, and the exact values of all the assumed parameters were simply selected to provide a by-eye fit to the data. Despite this, the functional form produced by such a toy model reasonably matches the behaviour seen in the data, suggesting that the suppression of star formation we observe may be driven by the fraction of gas which is inside the turnover radius. 

In panel b) of Figure \ref{tdeprotcurve} we plot the gas depletion time against the fraction of the molecular gas mass that lies within the turnover of the rotation curve $\left(\frac{M(r<R_{\rm peak})}{M_{\rm tot}}\right)$. We calculate this using the resolved CO maps from Paper XVIII, laying down an elliptical aperture at the turnover radius (with ellipticity calculated using the inclinations of these objects from Paper V; \citealt{2011MNRAS.414..968D}), and determining the fraction of the CO flux coming from inside this radius. Despite the limited resolution of our CO maps causing significant scatter, this panel broadly confirms our interpretation of panel a, showing that systems with the majority of their molecular gas mass lying within the turnover of the galaxy rotation curve have longer depletion times. {The dashed line is a guide to eye fitted to the black points, and has the form:}

\begin{equation}
 \mathrm{log}(T_{\rm dep})=(0.73\pm 0.11) \frac{M(r<R_{\rm peak})}{M_{\rm tot}} - (0.20\pm0.08).
\end{equation}

In panel c) of Figure \ref{tdeprotcurve} we plot the depletion time against the logarithmic derivative of the rotation curve ($\beta$; as defined in Section \ref{kdmsfr_rel}), a variable in defining the stability of the gas disc (see \citealt{2012ApJ...745...69K}). We average this quantity over the inner portion of the galaxy, where the rotation curve is rising, as this is the area that seems to be involved in star formation suppression (see panel a and b). This variable describes how steeply rising the rotation curve is, with values of zero being expected for a flat rotation curve, and values of 1 representing solid body rotation. Figure \ref{tdeprotcurve}, panel c) shows that these variables anti-correlate, although with reasonable scatter (Spearmans rank correlation coefficient of -0.6). On average it seems that galaxies with the strongest suppression of star formation have fast rising rotation curves, which plateau before reaching the peak velocity, leading to $\beta$ values as low as $\approx$0.1. Galaxies with more normal depletion times, on the other hand, have rotation curves that rise more steadily to a peak, with average values of $\beta \approx$0.35.

Following directly from panel c, it should be noted that the shear rate (A/$\Omega$) is directly related to the derivate of the galaxy rotation curve, as below:

\begin{equation}
\frac{A}{\Omega} = -0.5(\frac{1}{\Omega} \frac{dV}{dR}-1)
\end{equation}

\noindent where A is the first Oort constant, $\Omega$ is the angular velocity ($\propto$ V/R), V is the circular velocity, and R is the radius. Thus the depletion time would positively correlates with the mean shear rate in the inner regions.
Thus it seems that galaxies which do not form stars efficiently have higher shear rates than those with more normal star formation efficiencies.  It is not clear if shear can really be the only factor governing the SFE in ETGs however, as A/$\Omega$ is even higher in the flat part of these galaxies rotation curves, where the SFE appears to be normal.%

\section{Discussion}

\subsection{22~\mum\ emission from old stars}

In Figure \ref{WISEpassive} we explored the relationship between 22~\mum\ emission and $K_{\rm s}$-band luminosity in CO non-detected ETGs.

A strong relationship was confirmed between $K_{\rm s}$-band magnitude and 22~\mum\ emission (as presented in T09), however the amount of emission from CO non-detected galaxies can vary by almost half an order of magnitude between objects of the same mass. AGN activity and galaxies with young stellar populations  ($<$4 Gyr) contribute to this scatter, but they do not dominate it. This poses interesting questions about the nature of this emission, and of the scatter, in molecule poor galaxies. 

It has been suggested that this mid-infrared emission comes from circumstellar material around (post-) asymptotic giant branch (AGB) stars \citep{2002ApJ...571..272A}.
All stars between 0.8 and 8 $M_{\odot}$ will go through an AGB stage \citep{2005ARA&A..43..435H}, but from a single burst of star formation the number of stars going through this (relatively short-lived) phase is a strong function of time. Additionally, the dust production rate from AGB stars are thought to scale with luminosity, so higher mass AGB stars (that die quickly) produce more dust than their lower mass counterparts \citep{2012ApJ...748...40B}. As massive ETGs have harsh radiation fields, dust would be expected to be destroyed (through sputtering) on short timescales ($\ltsimeq$46 Myr; \citealt{2010A&A...518L..50C}) if no cold-ISM is available to shield it, so a constant supply of new dust is required.
Overall we would expect that galaxies that have not formed many new stars in the last 12 Gyr should not have as many AGB stars (per unit luminosity, or mass), and thus not as much 22~\mum\ emission as relatively young galaxies. 
As also found by \cite{2005ApJ...635L..25T}, there seems to be no relation between the amount of 22~\mum\ emitting dust and stellar population age beyond 4 Gyr, and hence perhaps another source of dust may be important in old galaxies.

Many other potential dust production mechanisms \citep[such as supernovae; e.g. ][]{2011Sci...333.1258M} should also be strongly linked to the stellar population age, and so cannot explain either the dust emission observed in these old galaxies, nor the residuals about the best-fit relation in Figure \ref{WISEpassive}. Mergers could bring in new dust (and new stars that produce dust), but they cannot explain the smooth distribution of this dust throughout these (gas poor) galaxies \citep{2002ApJ...571..272A}, and the strong link between the stellar luminosity of the galaxy and the warm dust emission. Mergers could potentially contribute to the scatter seen at fixed galaxy luminosity seen in Figure \ref{WISEpassive}, but given the short lifetime of dust in these objects, the merger rate would have to be high.

Perhaps emission from very small grains (VSGs; \citealt{1984ApJ...277..623S}) could help explain dust emission from these old passive galaxies. VSGs seem to have a longer lifetime in the ISM \citep{2010MNRAS.407L..49H}, and are produced during the destruction of larger crystalline dust grains. The intrinsic scatter around the best-fit relation in Figure \ref{WISEpassive} remains largely not understood, however a full exploration of this phenomenon is beyond the scope of this paper.

\subsection{Star formation rates}

For this sample of molecular gas-rich ETGs we find SFRs between $\approx$0.01 and 3 $M_{\odot}$ yr$^{-1}$, and SFR surface densities ranging from $\approx$0.004 to 18.75 $M_{\odot}$ yr$^{-1}$ kpc$^{-2}$. The median star-formation rate for our molecular gas-rich ETGs is $\approx$0.15 $M_{\odot}$ yr$^{-1}$, and the median star formation surface density is 0.06 $M_{\odot}$ yr$^{-1}$ kpc$^{-2}$. We find that almost all molecule rich ETGs have higher star-formation rate surface densities than average spiral galaxies. They lie in the same region of the Kennicutt-Schmidt plot  (Figure \ref{WISE22umfig}) as the spiral galaxy centres from K98. 
This may be a selection effect, as galaxies with widespread star formation would likely not have been morphologically classified as early type, and our flux-limited CO survey biases us to objects with high molecular gas surface brightness. A deeper survey would be required to determine if some ETGs have low surface density disks like those found in spirals, or if such objects are truly absent.

Overall it is clear that simply selecting ETGs (by either colour or morphology) is not a good way to ensure a galaxy sample is free from star formation activity, as is often assumed by studies at higher redshifts.

 \subsection{Star formation relations and efficiencies}
 \label{ks_discuss}
Figure \ref{WISE22umfig} clearly shows that ETGs, on average, form a factor of $\approx$2.5 fewer stars per unit molecular gas mass than late-type and starburst galaxies (and a factor $\approx$20 fewer than high-redshift starbursts). 

Our estimate of a lower SFE by a factor of $\approx$2.5 agrees well with the decrease of star formation efficiency observed for galaxies with redder colours, higher stellar mass concentrations, and/or higher stellar mass densities in the COLD-GASS survey \citep{2011MNRAS.415...61S,2012ApJ...758...73S}, and (for a subsample of 8 of the objects studied here) in the resolved star formation study published in Paper XXII.
This factor of two also is similar to predictions from simulations of gas in idealised galaxies which is affected by `morphological quenching' (\citealt{2009ApJ...707..250M} and Paper XXII).

It is clear that ETGs do not not fit well in a picture where the star-formation efficiency is assumed to be constant \citep[e.g.][]{2008AJ....136.2846B}, or even in a bimodal theory with star bursting and a regular star-formation modes. Instead, as we show above that the SFE varies smoothly as a function of galaxy properties, a likely more physical model would be that a continuum of star formation modes exist, spanning the range between extreme high SFE starbursts and our low efficiency early-type objects.

In our study, this difference is mainly driven by galaxies with star formation rates below $\approx$~0.3 $M_{\odot}$ yr$^{-1}$ kpc$^{-2}$ (or equivalently cold gas surface densities $<$~300 $M_{\odot}$ pc$^{-2}$). The few systems above this limit are consistent, within their errors, with following a standard Kennicutt-Schmidt relation (and lie above the constant SFE relation of B08 in Figure \ref{WISE22umfig_bigiel}). These systems generally have dense circumnuclear gas reservoirs, and would likely be considered as central starbursts if located in a late type galaxy. 

The systems with the lowest SFEs all have cold gas surface densities of around $\approx$~100 $M_{\odot}$ pc$^{-2}$. These systems tend to have extended molecular gas discs, that appear to be dynamically relaxed and follow exponential molecular gas surface brightness profiles (Paper XIV), but have the majority of their gas situated in the rising part of the galaxy rotation curve (See Section \ref{tdep_turnover}).
We below consider if the offset observed in these systems could be caused by changing gas properties, or the difficulty of estimating SFRs and molecular gas masses. 

\subsubsection{Difficulties in estimating star formation rates}
 
SFRs are notoriously difficult to calibrate. They rely intimately on knowing the number of massive stars formed in a given star formation episode, and thus the number of ionising photons. The SFR calibrations we used here have all been calibrated in normal star-forming spiral/starburst galaxies. However, the physical conditions within our early-type galaxies may be different in ways that violate the assumptions made in these calibrations. For instance, if the dust properties (e.g. size distribution or composition) were different in ETGs, then our 22~\mum\ fluxes could be systematically higher or lower than expected. 

The formation of dust is a controversial subject, but it is thought that stellar winds from AGB/pAGB stars are likely to be important \citep[e.g.][]{2006A&A...447..553F}. As ETGs tend to be metal rich and have large $\alpha$-element enhancements, the dust formed in the mass loss of such stars could be different from that found in late-type objects.  If dust is accreted from external sources, of course, then that complicates matters further (\citealt{2010A&A...518L..50C}, \citealt{2011MNRAS.417..882D}, hereafter Paper X). Additionally, all SFR calibrations make assumptions about the star-formation history of the objects (usually that the SFR has been constant over the past $\approx$100 Myr to 1 Gyr). In our early-type galaxies, where much of the gas may have been accreted recently, a much more bursty SF history may be more applicable. 

Recent evidence has suggested that the IMF is unlikely to be universal \citep[e.g.][]{2010Natur.468..940V,2012Natur.484..485C}, and varies as a function of galaxy properties. 
In particular in Paper XX we found the mass normalisation of the IMF to be related to the bulge fraction, which is also correlated to galaxy quenching. 
We have assumed here a Kroupa IMF for every object, but any object-to-object variation that depends on galaxy or ISM physical properties could affect the star-formation relation retrieved. Our objects are present in the sample of \cite{2012Natur.484..485C}, however, and we do not see any clear trend between the SFE and the IMF. 

Another factor that can affect our SFR determination is the inter-stellar radiation field (ISRF). In the spiral/starbursting systems where our SFR relations are calibrated, the ISRF is dominated by irradiation from newly formed OB stars. It is this light that we see directly in the FUV, and re-radiated in the infrared. In ETGs, however, various population of old stars generate intense hard radiation fields, that can dominate the ionisation structure of the ISM \citep{2010MNRAS.402.2187S}. The most massive ETG systems also host large X-ray halos, that provide an additional source of heating. We note however that in general these processes would increase the fluxes of the radiation we are using to trace star formation, and would thus make us overestimate the SFR in these galaxies. This can therefore not remove the discrepancy present in Figure \ref{WISE22umfig}, where our calculated star formation rates are low by a factor $\approx$2.5.

\subsubsection{A changing $X_{\rm CO}$}
\label{xcochange}
The molecular gas surface densities used here assume a Galactic $X_{\rm CO}$ factor (the conversion between CO flux and H$_2$). $X_{\rm CO}$ has been shown to vary as a function of metallicity \citep[e.g. ][]{1995ApJ...448L..97W,2008ApJ...686..948B,2011ApJ...737...12L,2012arXiv1212.1208S} and in high-redshift starbursts \citep[e.g.][]{1997ApJ...478..144S,1998ApJ...507..615D}. Other observational evidence from local starbursts \citep{2006ApJ...646..872H,2010AJ....140.1294M}, the Galactic centre \citep{1998ApJ...493..730O} and high-$z$ sub-millimetre galaxies \citep{2008ApJ...680..246T} suggests that $X_{\rm CO}$ may also be lower in regions of high molecular surface density. See \cite{2013ARA&A..51..207B} for a review of this issue. 

Our objects are generally metal rich (with metallicities around solar or above), where X$_{\rm CO}$ changes little. Thus including the effect of a varying $X_{\rm CO}$ with metallicity would change little the SFE of our ETGs.  
We also find no correlation between the SFE of our objects and stellar metallicity (although it is possible that the gas-phase metallicity is different, if gas has been accreted; Paper X).
The objects that show the largest SFE suppression in our data also do not have particularly dense molecular gas reservoirs, when compared to other systems where density driven $X_{\rm CO}$ variations have been observed.

As part of the BIMA-SONG survey, \cite{2001ApJ...561..218R} showed that molecular gas embedded in the bulges of spiral galaxies seems to emit more strongly (in the CO(1-0) line) than gas in the disc regions. \cite{2012arXiv1212.1208S} suggest that this may be due to a changing $X_{\rm CO}$, possibly caused by the higher ISM pressure within a large bulge. 
However, \cite{2013AJ....146...19L} found that SF in spiral galaxy centres is more efficient than normal even when taking into account a variable $X_{\rm CO}$, the opposite of the effect observed here.
By construction our systems are even more bulge-dominated than the objects studied by \cite{2013AJ....146...19L}, and thus the effects of a changing $X_{\rm CO}$ could be stronger. The galaxies with the strongest observed star formation suppression also have the majority of their gas in the central regions of the galaxy, where they would be most susceptible to this issue.

Overall we are unable to rule out the possibly that a lower $X_{\rm CO}$ in individual objects could be contributing to the scatter in the relations we observe. However, we find it unlikely that $X_{\rm CO}$ is lower in all objects uniformly, as then our ETGs would no longer lie on the KDM12 relation (as presented in Figure \ref{tffplot}) instead lying above it on average, forming more stars per unit mass and free-fall time. 

\subsubsection{Changing gas properties}

If the gas reservoirs in the objects that have low SFEs were to be substantially different from those found in other galaxies, this could help explain the discrepancy. For instance, if shear and/or bulge pressure increased the amount of low density CO emitting gas, this could cause us to overestimate molecular gas masses (see Section \ref{xcochange} above). Equally, if the fraction of dense gas were lower than normal in these objects, this could reduce the number of stars that are able to form. 

To search for signs of such changes in molecular gas properties, we correlated the SFE of our molecular gas rich ETGs with the molecular gas line ratios presented for some of these objects in \cite{2010MNRAS.407.2261K,2012MNRAS.421.1298C} and \cite{2013MNRAS.433.1659D}. Some of these objects with molecular line ratio information have low SFE, while others are consistent with having the same SFE as spiral galaxies. The $^{12}$CO/$^{13}$CO ratio is usually assumed to be an indicator of the mean optical depth of the CO emitting gas. If the amount of low density gas in these systems is increasing due to dynamical/pressure effects, one might expect the objects with low star formation efficiencies to have smaller mean optical depths. We find however that the $^{12}$CO/$^{13}$CO ratio does not correlate with the star formation efficiency in these objects, and (as shown by \citealt{2012MNRAS.421.1298C}) the $^{12}$CO/$^{13}$CO ratios in these objects are similar to those usually found in spiral galaxies. 

In a similar way, if the dense gas fraction is lower in objects with suppressed star formation then we would expect the ratio of dense gas tracers (such as HCN, HCO$^+$ and CS) to CO isotope emission to be suppressed.
 We find here that the HCN/CO, HCO$^+$/CO and CS/CO ratios do not correlate with the SFE, and again are in the normal range found for spiral galaxies. This suggests that the properties of the molecular clouds in these galaxies are not different in any systematic way that could explain the observed SFE suppression. 
  This is backed up by the analysis in \cite{2013MNRAS.432.1742B} and \cite{2013MNRAS.433.1659D}, who found the gas in these ETGs has similar physical properties (density, temperature) to gas in normal spiral galaxies and the Milky Way.

\subsection{Dynamically regulated star formation}
\label{dyn_sf_discuss}

In Section \ref{ks_discuss} we discussed the possibility that the low star formation efficiency we observe in these objects could be an artefact of changing gas properties, or the difficulty of estimating star formation rates and molecular gas masses. We find no evidence that definitively points to such a solution for this discrepancy. Indeed, one of the strongest arguments against such solutions are that they do not adequately explain the systematic behaviour of the star formation efficiency as a function of galaxy dynamics. In this section we discuss the possibility that these dynamical effects play a dominant role in regulating star formation in this sample of fast-rotating ETGs. 

We showed above that one can obtain a single star-formation relation that fits ETGs, low and high-redshift spiral/starburst galaxies and galactic clouds, if you normalise the gas density by the local free-fall time, as shown in Figure \ref{tffplot}.  
Our ETGs are constant with using up $\approx$1\% of their gas per local free-fall time, and the scatter around this relation is reasonably low. 

Despite this, Figure \ref{tffdeprotcurve} in Appendix C shows that once again, additional parameters correlate with the scatter seen around the KDM12 relation. The shape of the potential ($\beta$) and relative gas extent all correlate with the ratio of the depletion time to the free-fall time. If the gas velocity dispersion, or average cloud mass, which we have assumed to be constant, change systematically with these properties then this could lead to these dependancies. 

We briefly discussed above the possible importance of shear. Shear induced by galactic rotation acts to prevent gravitational collapse of gas clouds, which will increase the Jeans mass required for them to become bound, potential influencing the mean cloud density \citep{1964ApJ...139.1217T}. In addition, strong shear that pulls clouds apart, or an increased presence of hot gas in galaxy centres could increase gas velocity dispersions.  A similar correlation between shear and specific star formation rate has been found in the discs of some spiral galaxies \citep{2005MNRAS.361L..20S}, suggesting this process may be important (however, it should be mentioned that shear appears to matter little in Milky Way star-forming regions and some spiral objects \citealt{2012ApJ...758..125D,2013arXiv1304.7910M}).

KDM12 also present an alternative prescription for calculating the free-fall time, that assumes that star formation is regulated by the dynamical stability of a continuous disc of gas, which globally should have a Toomre Q parameter \citep{1964ApJ...139.1217T} of $\approx$1. In this case dynamical time and $\beta$ enter the equation for free-fall time directly. In Figure \ref{tffdeprotcurve} we do not use this formalism, as GMC timescales dominate, but the remaining correlations with these parameters suggest the global stability of the disc could still be important.

\section{Conclusions}

In this paper we presented star formation rates for the (fast-rotating) molecule-rich \atlas\ early-type galaxies, derived from \textit{WISE} 22~\mum\ and \textit{GALEX} FUV data.  
We first recalibrated the relation between $K_{\rm s}$-band luminosity and 22~\mum\ emission for our large sample of CO non-detected ETGs, to allow subtraction of 22~\mum\ emission from circumstellar material around old stars. 
The emission from CO non-detected galaxies can vary by almost half an order of magnitude between objects of the same stellar mass. AGN activity and galaxies with younger stellar populations  ($<$4~Gyr) contribute to this scatter, but do not dominate it. We were unable to reproduce claimed correlations between stellar metallicity and the scatter in this relation, and thus the astrophysical driver of the majority of the scatter remains unknown.
 
Once the contribution from old stars has been removed, we found SFRs between $\approx$0.01 and 3 $M_{\odot}$ yr$^{-1}$, and SFR surface densities ranging from $\approx$0.004 to 18.75 $M_{\odot}$ yr$^{-1}$ kpc$^{-2}$. The median SFR for our molecular gas-rich ETGs is $\approx$0.15 $M_{\odot}$ yr$^{-1}$, and the median SFR surface density is 0.06 $M_{\odot}$ yr$^{-1}$ kpc$^{-2}$.
Almost all molecule-rich ETGs have higher SFR surface densities than the disk of the average spiral galaxy, but similar to spiral galaxy centres.
This is depute many of the galaxies being bulge-dominated, and lying in the red-sequence on an optical colour magnitude diagram. It is thus clear that selecting early-type objects by morphology or optical colour is not a good way to build a sample free from star formation activity, as is often assumed by studies at higher redshifts. 

Using these SFRs, we showed that our ETGs fall below the canonical Kennicutt-Schmidt relation, forming on average a factor of $\approx$2.5 fewer stars per unit molecular gas mass than late-type and starburst galaxies (and a factor of $\approx$20 fewer than high-redshift starbursts). 
In our study, this difference is mainly driven by galaxies with SFRs below $\approx$0.3 $M_{\odot}$ yr$^{-1}$ kpc$^{-2}$ (or equivalently cold gas surface densities $<$300 $M_{\odot}$ pc$^{-2}$). These systems have the majority of their molecular gas concentrated in the inner regions of their host galaxy where the rotation curve is still rising, and shear is high. 

A local dynamical star-formation relation (taking into account the local free-fall time within the galaxy disc) reproduces well our observations. 
Using this relation one can obtain a single star-formation relation, that fits ETGs, Galactic clouds and spiral/starburst galaxies at all redshifts. Despite this, the residuals around the dynamical star-formation relation still correlate with galaxy properties such as the shape of the potential in the inner regions. We postulate that the dynamical stability of the gas may be an important second parameter, that suppresses star formation and causes much of the scatter around the best-fit dynamical star-formation relation.

We discussed various mechanisms that can cause this effect, and more generally the difficulties inherent in estimating SFRs and molecular gas masses in these ETGs. A changing X$_{\rm CO}$ factor could potentially cause the low SFE we observe, but it can not explain why the SFE in our study depends so strongly on dynamical quantities.

It is clear that further study will be required to fully determine the cause of the low SFE in ETGs. Ascertaining what is driving the residuals around the KS and KDM12 law will give us a direct way to probe the physics that regulates star formation. For instance, if variations in cloud properties and gas velocity dispersions are present in the central parts of ETGs, then they can potentially explain some of the SFE trends. Obtaining observational evidence for such variations will require high angular and spectral resolution observations, to resolve individual molecular clouds.  Gas-phase metallicity estimates and observations of multiple spectral lines could also be used to determine if the X$_{\rm CO}$ factor in these objects is systematically different. In addition, studies of the stability of the gas, and comparison with spatially-resolved star-formation relations, will be crucial to determine how changes in galactic conditions affect the physics of star formation.

\noindent \textbf{Acknowledgments}

The research leading to these results has received funding from the European
Community's Seventh Framework Programme (/FP7/2007-2013/) under grant agreement
No 229517. 
MC acknowledges support from a Royal Society University Research Fellowship.
This work was supported by the rolling grants ÔAstrophysics at OxfordÕ PP/E001114/1 and ST/H002456/1 and visitors grants PPA/V/S/2002/00553, PP/E001564/1 and ST/H504862/1 from the UK Research Councils. RLD acknowledges travel and computer grants from Christ Church, Oxford and support from the Royal Society in the form of a Wolfson Merit Award 502011.K502/jd. RLD is also grateful for support from the Australian Astronomical Observatory Distinguished Visitors programme, the ARC Centre of Excellence for All Sky Astrophysics, and the University of Sydney during a sabbatical visit.
SK acknowledges support from the Royal Society Joint Projects Grant JP0869822.
RMcD is supported by the Gemini Observatory, which is operated by the Association of Universities for Research in Astronomy, Inc., on behalf of the international Gemini partnership of Argentina, Australia, Brazil, Canada, Chile, the United Kingdom, and the United States of America.
TN and MBois acknowledge support from the DFG Cluster of Excellence `Origin and Structure of the Universe'.
MS acknowledges support from a STFC Advanced Fellowship ST/F009186/1.
PS is a NWO/Veni fellow.
MBois has received, during this research, funding from the European Research Council under the Advanced Grant Program Num 267399-Momentum.
LY acknowledges support from NSF grant AST-1109803.
The authors acknowledge financial support from ESO.

This paper is based on observations carried out with the IRAM Thirty Meter Telescope and the CARMA interferometer. IRAM is supported by INSU/CNRS (France), MPG (Germany) and IGN (Spain). Support for CARMA construction was derived from the states of California, Illinois, and Maryland, the James S. McDonnell Foundation, the Gordon and Betty Moore Foundation, the Kenneth T. and Eileen L. Norris Foundation, the University of Chicago, the Associates of the California Institute of Technology, and the National Science Foundation. Ongoing CARMA development and operations are supported by the National Science Foundation under a cooperative agreement, and by the CARMA partner universities. This publication also makes use of data products from the Wide-field Infrared Survey Explorer, which is a joint project of the University of California, Los Angeles, and the Jet Propulsion Laboratory/California Institute of Technology, funded by the National Aeronautics and Space Administration.

\bsp
\bibliographystyle{mn2e}
\bibliography{A3D_SFpaper}
\bibdata{A3D_SFpaper}
\bibstyle{mn2e}

\label{lastpage}

\appendix

\section{Passive sample}

\begin{table}
\caption{Properties of the CO non-detected ETG sample used in this work.}
\begin{tabular*}{0.45\textwidth}{@{\extracolsep{\fill}}l r r r}
\hline
Galaxy & L$_{Ks}$ & L$_{\rm 22\mu m,obs}$ & Outlier  \\
  & log(L$_{\odot}$) & log(ergs s$^{-1}$) & \\
 (1) & (2) & (3) & (4)  \\
 \hline
    IC0560   &      10.15   &      41.39   & x   \\
    IC0598   &      10.35   &      41.27   & x   \\
    IC3631   &      10.12   &      40.90   &  -  \\
   NGC0448   &      10.52   &      40.93   &  -  \\
   NGC0474   &      10.88   &      41.05   &  -  \\
   NGC0502   &      10.53   &      40.77   &  -  \\
   NGC0516   &      10.20   &      40.62   &  -  \\
   NGC0525   &      10.06   &      40.81   &  -  \\
   NGC0661   &      10.59   &      40.93   &  -  \\
   NGC0680   &      10.98   &      41.37   &  -  \\
   NGC0821   &      10.91   &      41.34   &  -  \\
   NGC0936   &      11.25   &      41.66   &  -  \\
   NGC1023   &      10.92   &      41.36   &  -  \\
   NGC1121   &      10.39   &      40.63   &  -  \\
   NGC1248   &      10.47   &      40.99   &  -  \\
   NGC1289   &      10.70   &      41.28   &  -  \\
   NGC2481   &      10.66   &      41.40   &  -  \\
   NGC2549   &      10.28   &      40.91   &  -  \\
   NGC2577   &      10.68   &      41.32   &  -  \\
   NGC2592   &      10.46   &      40.76   &  -  \\
   NGC2679   &      10.44   &      41.00   &  -  \\
   NGC2695   &      10.77   &      41.11   &  -  \\          
   NGC2698   &      10.64   &      41.15   &  -  \\          
   NGC2699   &      10.40   &      40.80   &  -  \\          
   NGC2778   &      10.20   &      40.58   &  -  \\          
   NGC2852   &      10.18   &      40.61   &  -  \\          
   NGC2859   &      10.96   &      41.32   &  -  \\          
   NGC2880   &      10.50   &      40.70   &  -  \\          
   NGC2950   &      10.48   &      41.01   &  -  \\          
   NGC2962   &      10.92   &      41.60   &  -  \\          
   NGC2974   &      10.76   &      41.83   & x   \\          
   NGC3098   &      10.40   &      40.99   &  -  \\          
   NGC3193   &      11.16   &      41.39   &  -  \\          
   NGC3226   &      10.61   &      41.45   & x   \\          
   NGC3230   &      10.98   &      41.39   &  -  \\          
   NGC3248   &      10.28   &      40.36   &  -  \\          
   NGC3301   &      10.62   &      41.54   & x   \\          
   NGC3377   &      10.42   &      40.90   &  -  \\          
   NGC3379   &      10.83   &      41.17   &  -  \\          
   NGC3384   &      10.72   &      41.06   &  -  \\          
   NGC3412   &      10.33   &      40.48   &  -  \\          
   NGC3414   &      10.90   &      41.48   &  -  \\          
   NGC3457   &      10.07   &      40.42   &  -  \\          
   NGC3458   &      10.56   &      40.95   &  -  \\          
   NGC3499   &      10.06   &      40.90   & x   \\          
   NGC3530   &      10.11   &      40.62   &  -  \\          
   NGC3605   &      10.04   &      40.52   &  -  \\          
   NGC3608   &      10.77   &      40.99   &  -  \\          
   NGC3610   &      10.79   &      41.28   &  -  \\          
   NGC3613   &      11.02   &      41.06   &  -  \\          
   NGC3630   &      10.58   &      40.96   &  -  \\          
   NGC3640   &      11.15   &      41.73   &  -  \\          
   NGC3641   &      10.05   &      40.38   &  -  \\          
   NGC3648   &      10.54   &      41.03   &  -  \\          
   NGC3658   &      10.69   &      41.46   &  -  \\          
     \hline
 \end{tabular*}
\parbox[t]{0.45 \textwidth}{ \textit{Notes:} Column one lists the name of the galaxy. Column 2 contains the $K_{\rm s}$ band luminosity of the galaxy, calculated using the 2MASS total $K_{\rm s}$-band magnitude and the distance to these objects as in Paper I, and assuming that the absolute magnitude of the Sun at $K_{\rm s}$-band is 3.28 mag \citep{Binney:1998p3454}. Column 3 contains the \textit{WISE} 22~\mum\ luminosity of the galaxy, calculated as described in Section \ref{data}, once again using the distances from Paper I. Column 4 lists galaxies that were flagged as outliers in our survival analysis fit (marked with an `x')  . These objects are likely to have a molecular ISM and star-formation but were not detected in CO, probably due to the fixed flux limit of our survey. }
\label{passivetable}
\end{table}
   \begin{table}
\contcaption{}{}
\begin{tabular*}{0.45\textwidth}{@{\extracolsep{\fill}}l r r r}
\hline
Galaxy & L$_{Ks}$ & L$_{\rm 22\mu m,obs}$ & Outlier   \\
  & log(L$_{\odot}$) & log(ergs s$^{-1}$) & \\
 (1) & (2) & (3) & (4) \\
 \hline  NGC3674   &      10.60   &      40.80   &  -  \\          
   NGC3694   &      10.25   &      41.90   & x   \\          
   NGC3757   &      10.17   &      40.38   &  -  \\          
   NGC3796   &      10.05   &      40.68   &  -  \\          
   NGC3941   &      10.54   &      41.02   &  -  \\          
   NGC3945   &      11.04   &      41.53   &  -  \\          
   NGC3998   &      10.64   &      41.74   & x   \\          
   NGC4026   &      10.52   &      40.94   &  -  \\          
   NGC4078   &      10.51   &      41.18   &  -  \\          
   NGC4143   &      10.55   &      41.27   &  -  \\          
   NGC4168   &      10.92   &      41.41   &  -  \\          
   NGC4179   &      10.58   &      40.68   &  -  \\          
   NGC4191   &      10.55   &      41.06   &  -  \\          
   NGC4215   &      10.68   &      41.06   &  -  \\          
   NGC4233   &      10.86   &      41.49   &  -  \\          
   NGC4251   &      10.78   &      41.36   &  -  \\          
   NGC4255   &      10.51   &      40.88   &  -  \\          
   NGC4261   &      11.38   &      42.01   &  -  \\          
   NGC4262   &      10.35   &      40.71   &  -  \\          
   NGC4264   &      10.51   &      40.96   &  -  \\          
   NGC4267   &      10.58   &      40.64   &  -  \\          
   NGC4278   &      10.83   &      41.38   &  -  \\          
   NGC4281   &      10.92   &      41.82   & x   \\          
   NGC4339   &      10.31   &      40.35   &  -  \\          
   NGC4340   &      10.52   &      40.87   &  -  \\          
   NGC4342   &      10.14   &      40.54   &  -  \\          
   NGC4346   &      10.33   &      40.77   &  -  \\          
   NGC4350   &      10.56   &      41.13   &  -  \\          
   NGC4365   &      11.40   &      41.49   &  -  \\          
   NGC4371   &      10.69   &      41.18   &  -  \\          
   NGC4374   &      11.36   &      41.81   &  -  \\          
   NGC4377   &      10.28   &      41.26   & x   \\          
   NGC4379   &      10.21   &      40.52   &  -  \\          
   NGC4382   &      11.36   &      41.78   &  -  \\          
   NGC4387   &      10.16   &      40.51   &  -  \\          
   NGC4406   &      11.33   &      41.73   &  -  \\          
   NGC4417   &      10.46   &      41.03   &  -  \\          
   NGC4434   &      10.33   &      40.80   &  -  \\          
   NGC4442   &      10.76   &      41.29   &  -  \\          
   NGC4458   &      10.02   &      40.26   &  -  \\          
   NGC4461   &      10.54   &      40.77   &  -  \\          
   NGC4472   &      11.62   &      41.85   &  -  \\          
   NGC4473   &      10.82   &      41.15   &  -  \\          
   NGC4474   &      10.22   &      40.87   &  -  \\          
   NGC4478   &      10.43   &      40.90   &  -  \\          
   NGC4483   &      10.05   &      40.37   &  -  \\          
   NGC4486   &      11.46   &      42.07   &  -  \\          
  NGC4486A   &      10.04   &      40.55   &  -  \\          
   NGC4489   &       9.95   &      40.13   &  -  \\          
   NGC4494   &      10.96   &      41.31   &  -  \\          
   NGC4503   &      10.60   &      40.90   &  -  \\          
   NGC4521   &      10.88   &      41.55   &  -  \\          
   NGC4528   &      10.13   &      40.14   &  -  \\          
   NGC4546   &      10.63   &      41.25   &  -  \\          
   NGC4550   &      10.22   &      40.60   &  -  \\          
   NGC4551   &      10.18   &      40.35   &  -  \\          
   NGC4552   &      11.03   &      41.37   &  -  \\          
   NGC4564   &      10.54   &      41.19   &  -  \\          
   NGC4570   &      10.70   &      41.35   &  -  \\          
   NGC4578   &      10.38   &      41.01   &  -  \\          
   NGC4608   &      10.49   &      41.00   &  -  \\          
   NGC4612   &      10.33   &      40.67   &  -  \\          
   NGC4621   &      10.97   &      41.29   &  -  \\          
   NGC4623   &      10.01   &      40.87   & x   \\          
       \hline
 \end{tabular*}
\end{table}
   \begin{table}
\contcaption{}{}
\begin{tabular*}{0.45\textwidth}{@{\extracolsep{\fill}}l r r r}
\hline
Galaxy & L$_{Ks}$ & L$_{\rm 22\mu m,obs}$ & Outlier   \\
  & log(L$_{\odot}$) & log(ergs s$^{-1}$) & \\
 (1) & (2) & (3) & (4)  \\
 \hline
  NGC4624   &      10.78   &      40.54   &  -  \\          
   NGC4636   &      11.06   &      41.65   &  -  \\          
   NGC4638   &      10.52   &      40.87   &  -  \\          
   NGC4649   &      11.50   &      42.38   & x   \\          
   NGC4660   &      10.39   &      40.92   &  -  \\          
   NGC4690   &      10.50   &      41.12   &  -  \\          
   NGC4697   &      10.88   &      41.20   &  -  \\          
   NGC4754   &      10.77   &      41.18   &  -  \\          
   NGC4762   &      11.10   &      41.79   &  -  \\          
   NGC4803   &      10.22   &      40.71   &  -  \\          
   NGC5103   &      10.26   &      40.77   &  -  \\          
   NGC5198   &      10.95   &      41.24   &  -  \\          
   NGC5308   &      10.96   &      41.09   &  -  \\          
   NGC5322   &      11.42   &      41.96   &  -  \\          
   NGC5353   &      11.36   &      41.96   &  -  \\          
   NGC5355   &      10.27   &      41.15   & x   \\          
   NGC5422   &      10.79   &      41.06   &  -  \\          
   NGC5473   &      11.01   &      41.52   &  -  \\          
   NGC5475   &      10.46   &      41.19   &  -  \\          
   NGC5485   &      10.76   &      41.36   &  -  \\          
   NGC5493   &      11.11   &      41.49   &  -  \\          
   NGC5500   &      10.08   &      40.57   &  -  \\          
   NGC5507   &      10.59   &      40.97   &  -  \\          
   NGC5557   &      11.26   &      41.49   &  -  \\          
   NGC5574   &      10.23   &      40.95   &  -  \\          
   NGC5576   &      10.97   &      41.17   &  -  \\          
   NGC5582   &      10.62   &      40.97   &  -  \\          
   NGC5611   &      10.19   &      40.55   &  -  \\          
   NGC5631   &      10.79   &      41.47   &  -  \\          
   NGC5638   &      10.83   &      41.24   &  -  \\          
   NGC5687   &      10.60   &      41.03   &  -  \\          
   NGC5770   &      10.17   &      40.50   &  -  \\          
   NGC5813   &      11.35   &      41.79   &  -  \\          
   NGC5831   &      10.79   &      41.12   &  -  \\          
   NGC5838   &      10.96   &      41.65   &  -  \\          
   NGC5839   &      10.32   &      40.68   &  -  \\          
   NGC5845   &      10.48   &      40.99   &  -  \\          
   NGC5846   &      11.32   &      41.58   &  -  \\          
   NGC5854   &      10.63   &      41.25   &  -  \\          
   NGC5864   &      10.76   &      41.18   &  -  \\          
   NGC5869   &      10.62   &      40.97   &  -  \\          
   NGC6010   &      10.72   &      41.23   &  -  \\          
   NGC6017   &      10.32   &      41.44   & x   \\          
   NGC6149   &      10.35   &      41.02   &  -  \\          
   NGC6278   &      10.99   &      41.43   &  -  \\          
   NGC6547   &      10.75   &      41.03   &  -  \\          
   NGC6548   &      10.59   &      42.04   & x   \\          
   NGC6703   &      10.85   &      41.23   &  -  \\          
   NGC7280   &      10.44   &      40.98   &  -  \\          
   NGC7332   &      10.81   &      41.38   &  -  \\          
   NGC7457   &      10.26   &      40.68   &  -  \\          
 PGC016060   &      10.37   &      41.73   & x   \\          
 PGC042549   &      10.40   &      41.36   & x   \\          
 PGC051753   &      10.08   &      40.53   &  -  \\          
 PGC054452   &       9.95   &      40.31   &  -  \\          
  UGC04551   &      10.48   &      40.54   &  -  \\          
  UGC06062   &      10.44   &      41.12   &  -  \\  
 \hline
 \end{tabular*}
\end{table}
\clearpage

\section{Constant star formation efficiency star-formation relations}
\label{sfeconstplots}
   \begin{figure*}
\begin{center}
\includegraphics[height=5.45cm,angle=0,clip,trim=0.0cm 1.6cm 0cm 0cm]{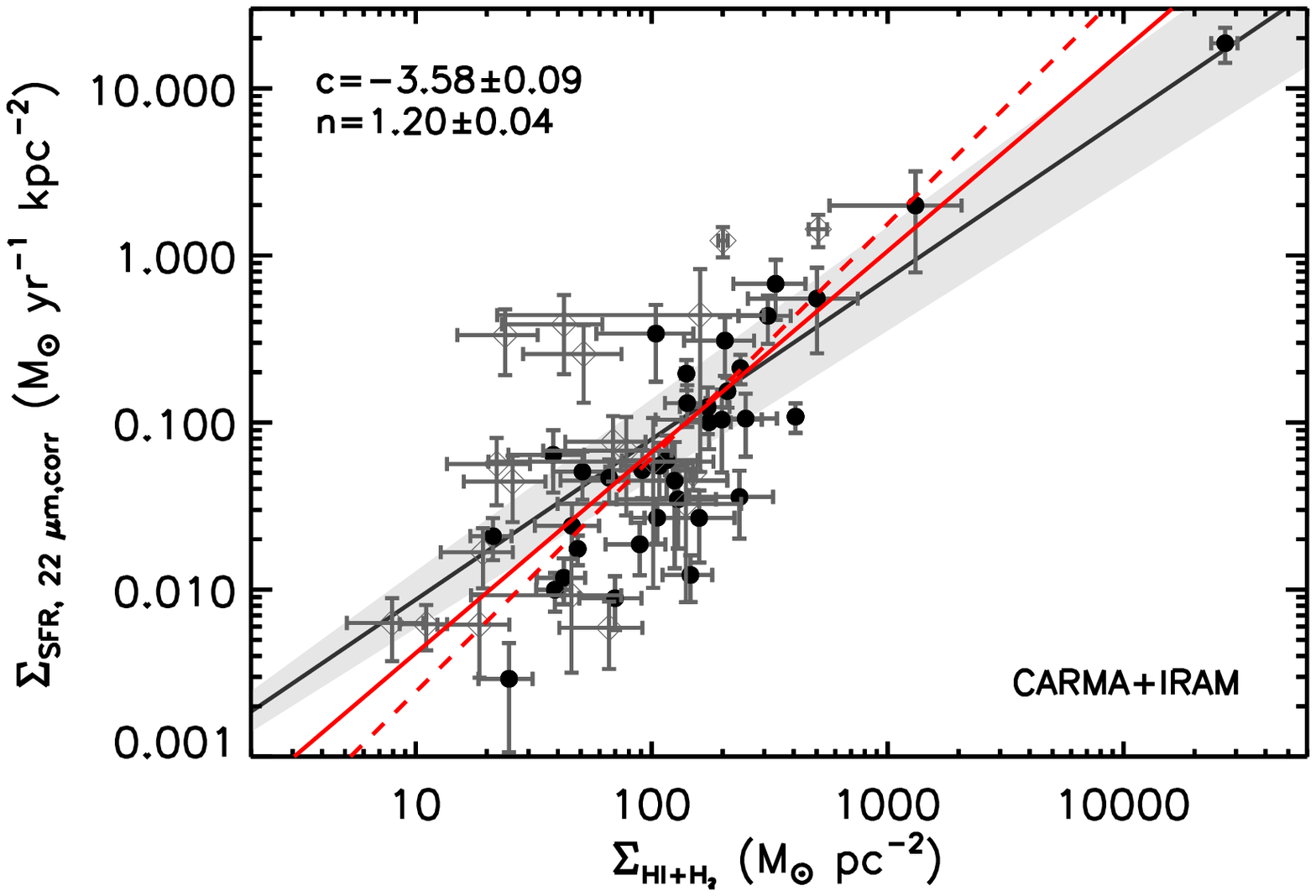}
\includegraphics[height=5.45cm,angle=0,clip,trim=3.1cm 1.6cm 0cm 0cm]{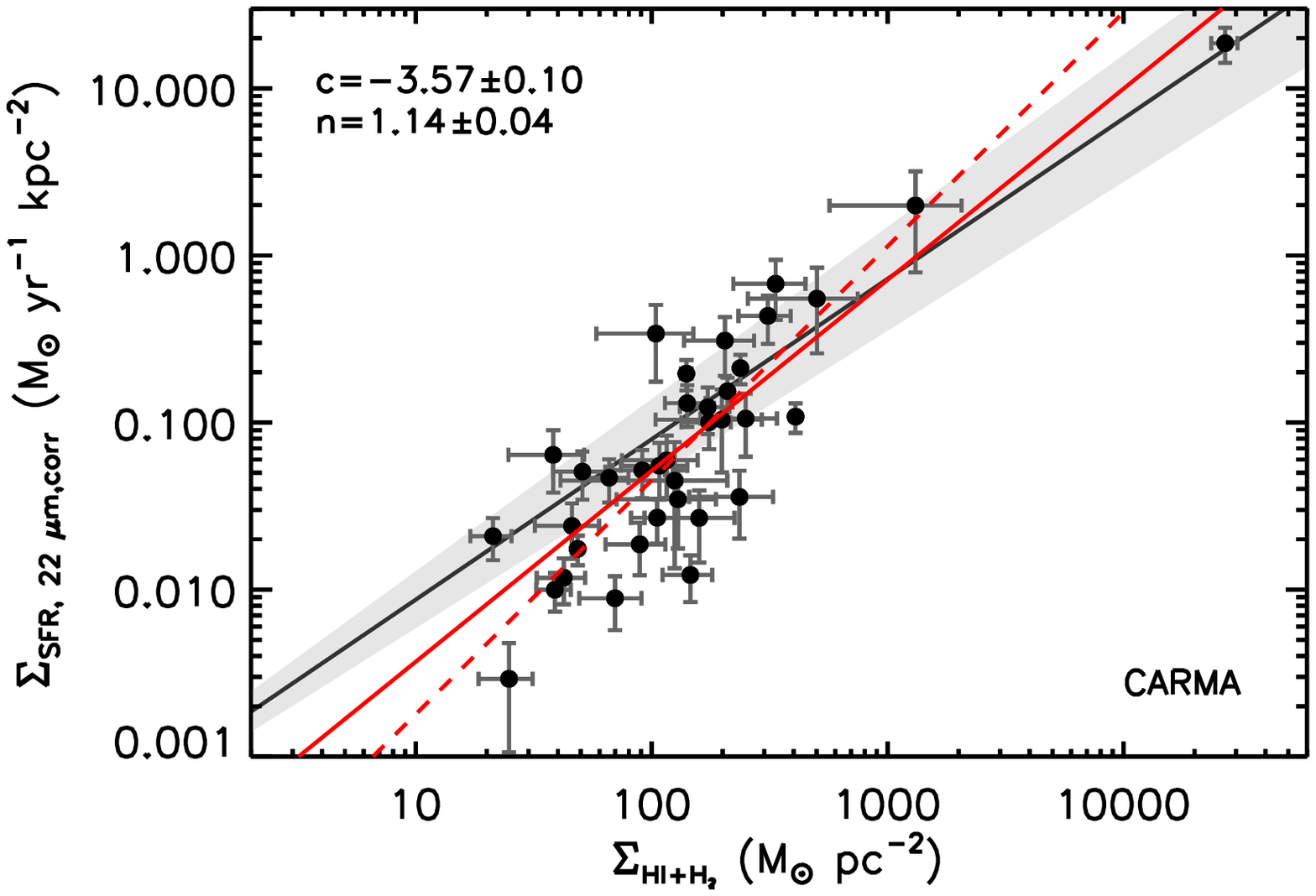}\\
\includegraphics[height=6.35cm,angle=0,clip,trim=0.0cm 0.0cm 0cm 0.0cm]{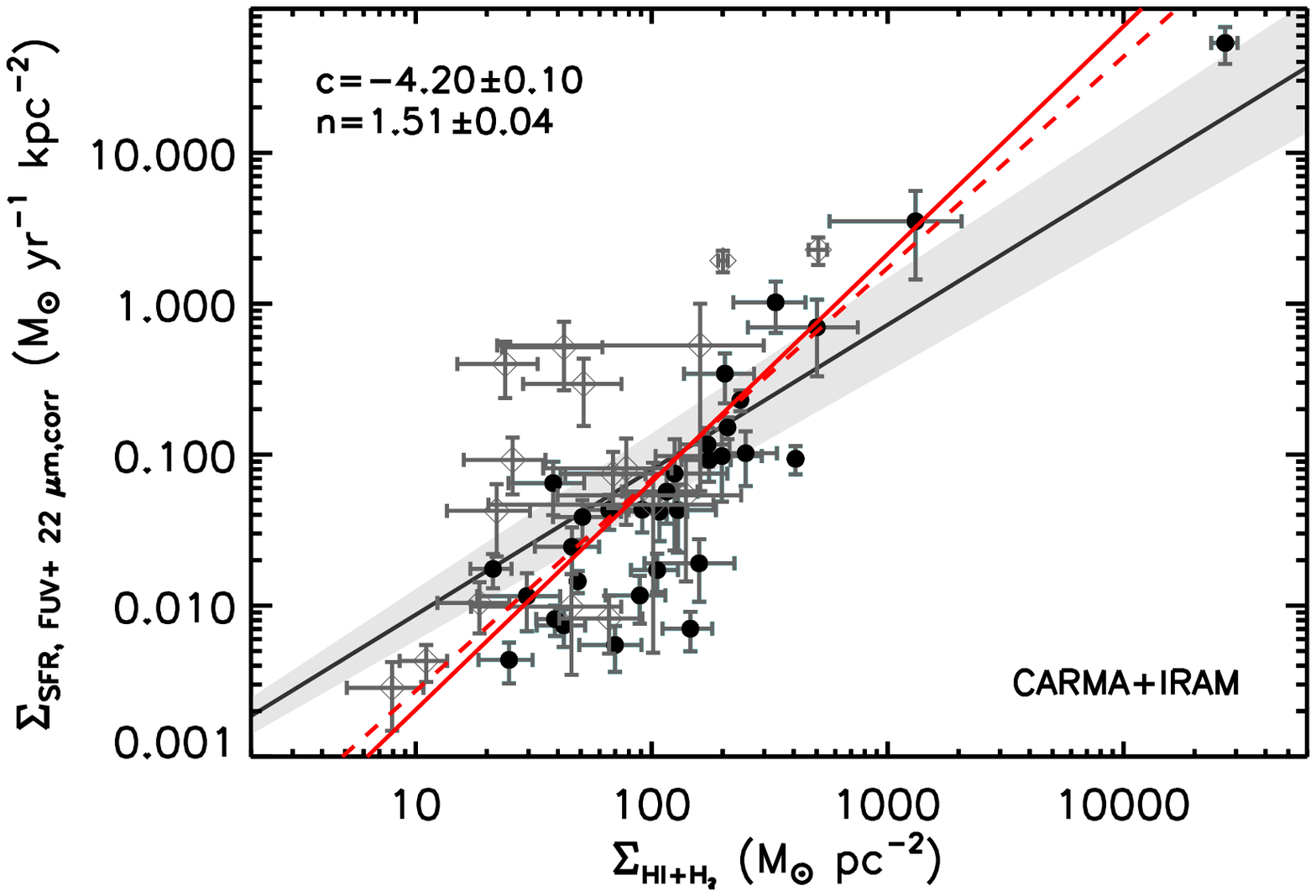}
\includegraphics[height=6.35cm,angle=0,clip,trim=3.1cm 0.0cm 0cm 0cm]{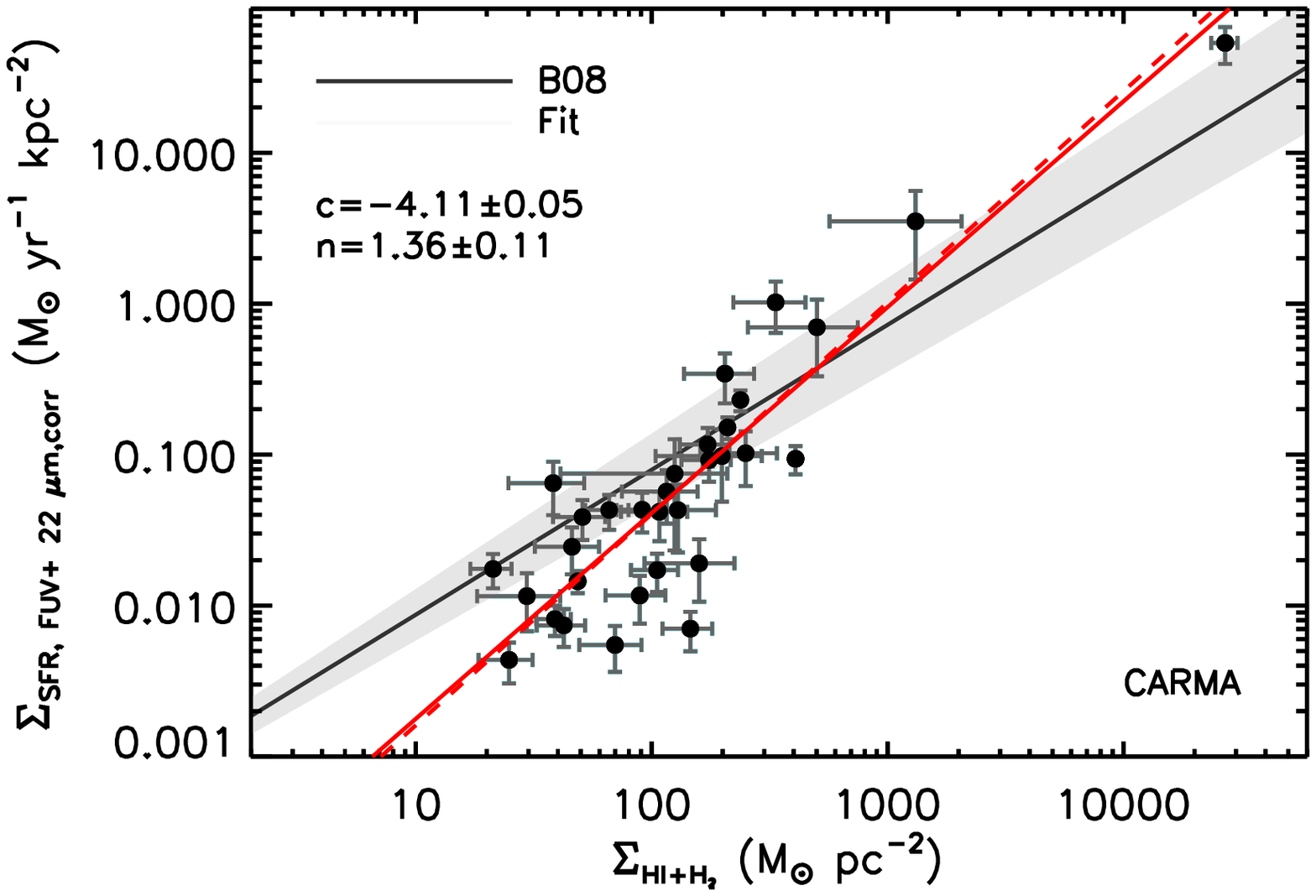}
 \end{center}
 \caption{As Figure \ref{WISE22umfig}, but showing the constant SFE relation of B08 as a black line, and its typical scatter as the grey shaded area.}
 \label{WISE22umfig_bigiel}
 \end{figure*}
\section{Residuals around dynamical star-formation relations}

\begin{figure}
\begin{center}
\subfigure{\includegraphics[height=6cm,angle=0,clip,trim=0.0cm 0cm 0cm 0cm]{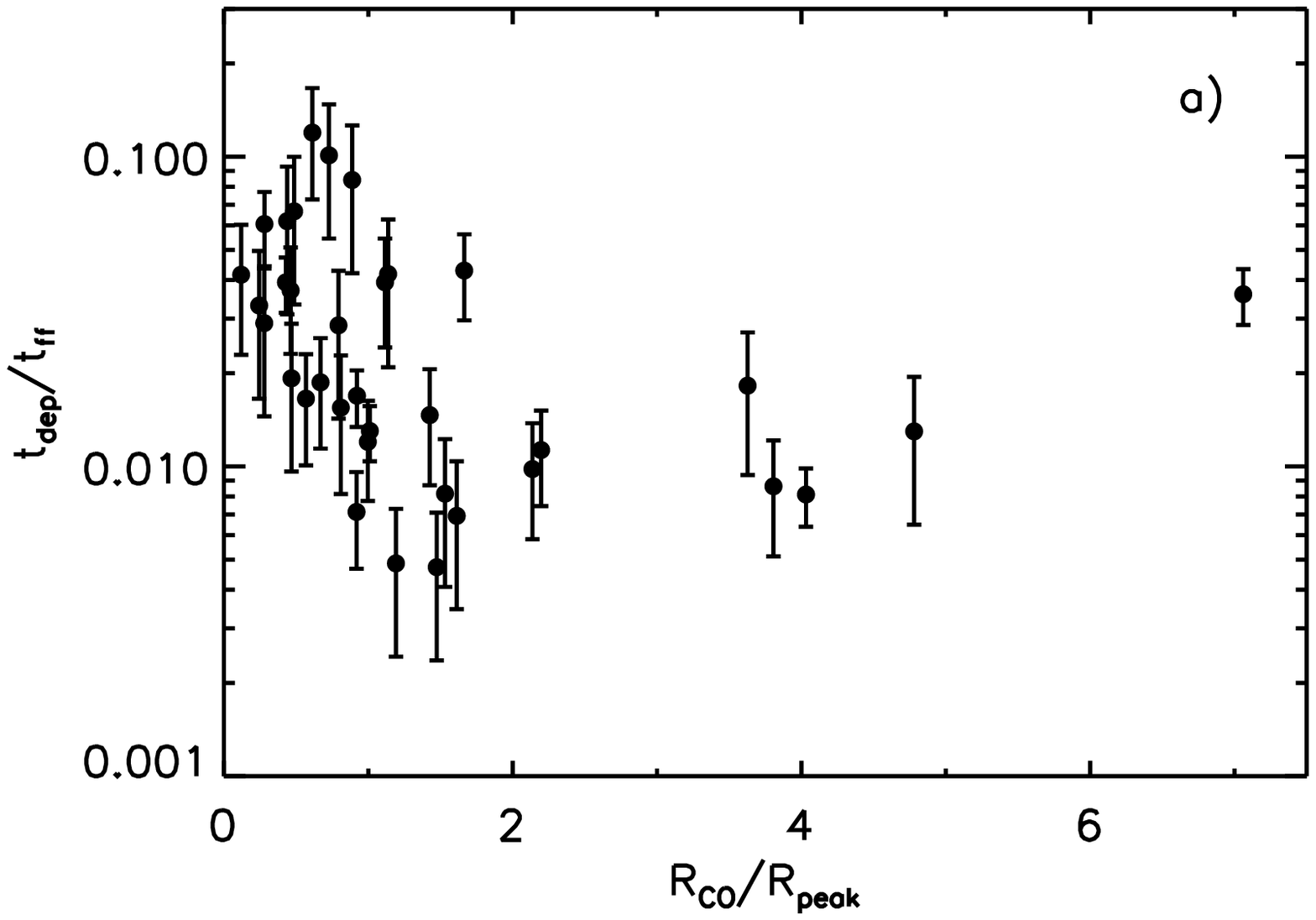}}
\subfigure{\includegraphics[height=6cm,angle=0,clip,trim=0.0cm 0cm 0cm 0cm]{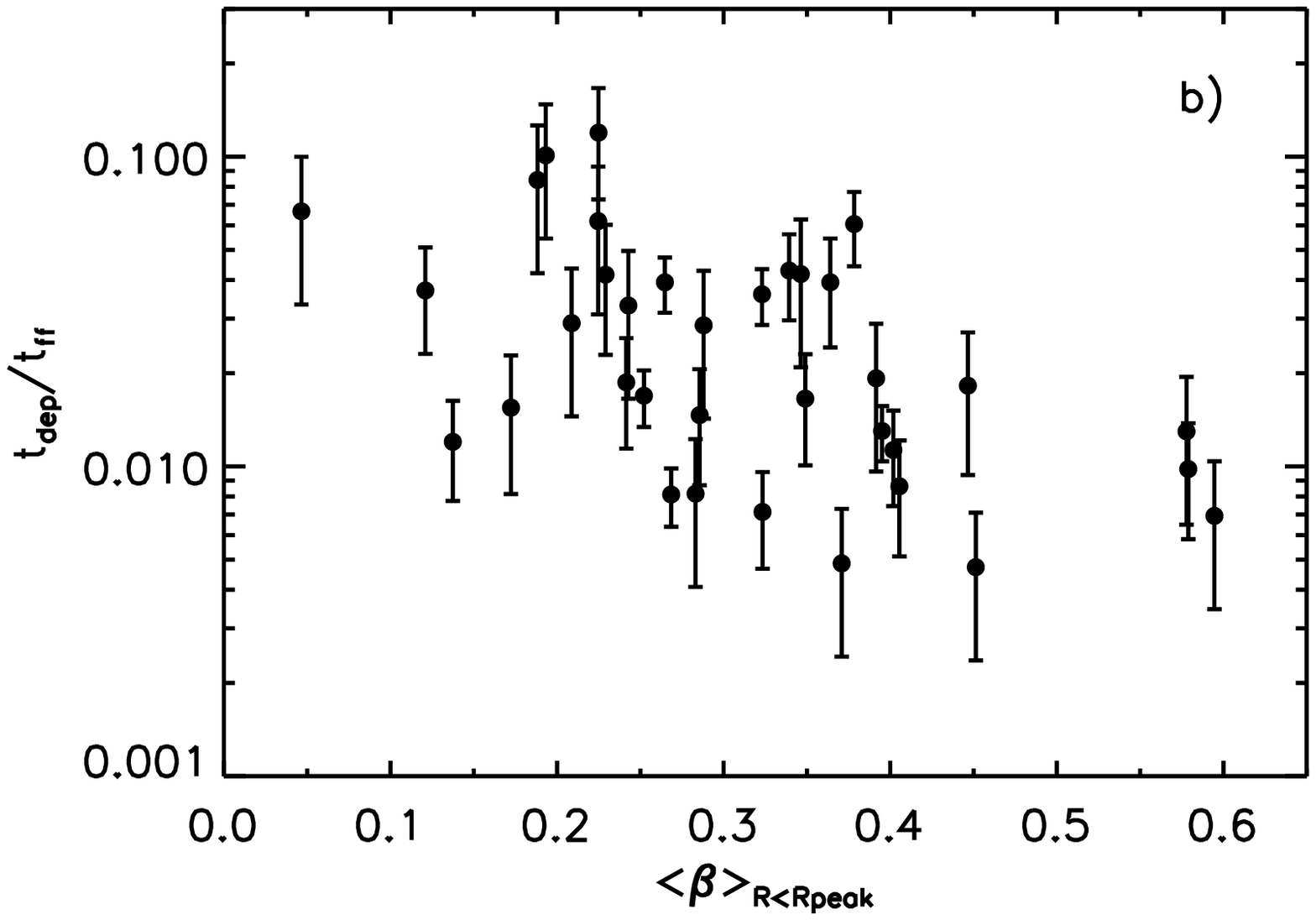}}
 \end{center}
 \caption{As in Figure \protect \ref{tdeprotcurve} (panels a \& c), but with gas depletion times normalised by the local free-fall time, calculated as described in \protect \cite{2012ApJ...745...69K}. This shows that residual dependancies on galaxy dynamics remain when normalising by the free-fall time alone.}
 \label{tffdeprotcurve}
 \end{figure}

\end{document}